\documentclass[english,useAMS,usenatbib]{mn2e}
\usepackage[T1]{fontenc}
\usepackage[utf8]{inputenc}
\usepackage{siunitx}
\usepackage{rotating}
\usepackage{url}
\usepackage{amsmath}
\usepackage{physics}
\usepackage{commath}
\usepackage{amssymb}
\usepackage{graphicx}
\usepackage{subfig}
\usepackage{multirow}
\usepackage[table]{xcolor}
\usepackage{pdfpages}
\usepackage{esint}
\usepackage[authoryear]{natbib}
\usepackage{listings}
\usepackage{textcomp}
\usepackage[export]{adjustbox}
\usepackage{booktabs}
\usepackage{ulem}
\usepackage{comment}

\makeatletter

\def\gtsima{$\, \buildrel > \over \sim \,$}
\def\ltsima{$\, \buildrel < \over \sim \,$}
\def\prosima{$\, \buildrel \propto \over \sim \,$}
\def\gsim{\lower.5ex\hbox{\gtsima}}
\def\lsim{\lower.5ex\hbox{\ltsima}}
\def\simgt{\lower.5ex\hbox{\gtsima}}
\def\simlt{\lower.5ex\hbox{\ltsima}}
\def\simpr{\lower.5ex\hbox{\prosima}}
\def\la{\lsim}
\def\ga{\gsim}

\newcommand{\be}{\begin{eqnarray}}
\newcommand{\ee}{\end{eqnarray}}
\def\lsim{\,\lower2truept\hbox{${< \atop\hbox{\raise4truept\hbox{$\sim$}}}$}\,}
\def\gsim{\,\lower2truept\hbox{${> \atop\hbox{\raise4truept\hbox{$\sim$}}}$}\,}

\providecommand{\tabularnewline}{\\}

\title[Pop\ III star forming environments]{A needle in a haystack? Catching Pop\ III stars in the Epoch of Reionization: I. Pop\ III star forming environments}
\author[Venditti et al.]{Alessandra Venditti$^{1,2,3,4}$\thanks{E-mail:alessandra.venditti@inaf.it}; Luca Graziani$^{1,2,3}$; Raffaella Schneider$^{1,2,3}$;  \newauthor Laura Pentericci$^{3}$;
 Claudia Di Cesare$^{1,2,3}$; Umberto Maio$^{5}$; Kazuyuki Omukai$^{6}$ \\
$^{1}$Dipartimento di Fisica, Sapienza, Universit$\grave{a}$ di Roma, Piazzale Aldo Moro 5, 00185, Roma, Italy\\
$^{2}$INFN, Sezione di Roma I, Piazzale Aldo Moro 2, 00185, Roma, Italy\\
$^{3}$INAF-Osservatorio Astronomico di Roma, Via di Frascati 33, 00078, Monte Porzio Catone, Italy\\
$^{4}$Dipartimento di Fisica, Tor Vergata, Universit$\grave{a}$ di Roma, Via Cracovia 50, 00133, Roma, Italy\\
$^{5}$INAF-Osservatorio Astronomico di Trieste, via G. Tiepolo 11, 34143, Trieste, Italy \\
$^{6}$Astronomical Institute, Graduate School of Science, Tohoku University, Aoba, Sendai 980-8578, Japan
}

\begin{document}

\date{\today}

\pagerange{\pageref{firstpage}--\pageref{lastpage}} \pubyear{2022}

\maketitle

\label{firstpage}

\begin{abstract}
Despite extensive search efforts, direct observations of the first (Pop\ III) stars have not yet succeeded. Theoretical studies have suggested that late Pop\ III star formation is still possible in pristine clouds of high-mass galaxies, coexisting with Pop\ II stars, down to the Epoch of Reionization (EoR). Here we reassess this finding by exploring Pop\ III star formation in six $50h^{-1} ~ \si{cMpc}$ simulations performed with the hydrodynamical code \texttt{dustyGadget}. We find that Pop\ III star formation ($\sim 10^{-3.4} - 10^{-3.2} ~ \si{M_\odot.yr^{-1}.cMpc^{-3}}$) is still occurring down to $z \sim 6 - 8$, i.e. well within the reach of deep JWST surveys. At these epochs, $\gtrsim 10 \%$ of the rare massive galaxies with $M_\star \gtrsim 3 \times 10^9 ~ \si{M_\odot}$ are found to host Pop\ III stars, although with a Pop\ III/Pop\ II mass fraction $\lesssim 0.1 \%$. 
Regardless of their mass, Pop\ III hosting galaxies are mainly found on the main sequence, at high star formation rates, probably induced by accretion of pristine gas. This scenario is also supported by their increasing star formation histories and their preferential location in high-density regions of the cosmic web. Pop\ III stars are found both in the outskirts of metal-enriched regions and in isolated, pristine clouds. In the latter case, their signal may be less contaminated by Pop\ IIs, although its detectability will strongly depend on the specific line-of-sight to the source, due to the complex morphology of the host galaxy and its highly inhomogeneous dust distribution.
\end{abstract}


\section{Introduction}
\label{sec:introduction}

The unprecedented sensitivity and resolution of JWST\footnote{\url{https://jwst.nasa.gov}} open up concrete perspectives in constraining the early stages of cosmic star formation by detecting observational signatures of the first generation of stars, either at Cosmic Dawn or in galaxies evolving in the EoR ($z \gtrsim 6$).

At Cosmic Dawn, a complementary probe of the first sources of light is encoded in the cosmological 21-cm signal of neutral hydrogen (see e.g. \citealt{Madau_1997, Ciardi_2003b, Mesinger_2011, Visbal_2012, Fialkov_2013, Fialkov_2014} for some fundamental contributions). So far, the only claimed detection of the 21-cm signal has been reported by the global signal experiment EDGES \citep{Bowman_2018}. Although still debated (see e.g. \citealt{Hills_2018, Bowman_2018_reply, Singh_Subrahmanyan_2019, Bradley_2019, Sims_Pober_2020, Tauscher_2020, Singh_2022}), the signal, centered  $z \sim 17$, suggests an early start of star formation and shows the amazing potential of 21-cm observations to constrain the nature of the first stars \citep{Jana_2019, Schauer_2019, Fialkov_Barkana_2019, Reis_2020, Mebane_2020, Chatterjee_2020, Gessey-Jones_2022}, the properties of the first accreting black holes \citep{Ewall-Wice_2018, Ewall-Wice_2019, Mirabel_2019, Ventura_2023}, and the mode of star formation in low-mass dark matter halos \citep{Mirocha_Furlanetto_2019}.

In the EoR, a supposedly individual star or small stellar multiple named ``Earendel'' has been observed by the Hubble Space Telescope (HST) at $z = 6.2$, thanks to gravitational lensing \citep{Welch_2022}, and many more extremely compact systems are expected to be seen at high redshifts by JWST. Six young, massive star clusters with measured radii spanning $\sim 20 ~ \si{pc}$ down to $\sim 1 ~ \si{pc}$ have been found by \citealt{Vanzella_2023} in the same ``Sunrise Arc'' galaxy hosting Earendel. Other objects of this kind have also been found - although at a lower redshift, $z = 1.378$ - in the JWST/NIRCam images of a gravitationally lensed field around the ``Sparkler'' galaxy \citep{Mowla_2022_JWSTclusters, Lee_2022_JWSTclusters, Faisst_2022_JWSTclusters}; interestingly enough, a significant number of these sources appear unresolved in the NIRCam/F150W filter, suggesting remarkably small sizes ($< 50 ~ \si{pc}$), that are consistent with local globular clusters and even lower than local star-forming clumps or dwarf galaxies \citep{Faisst_2022_JWSTclusters}. While Pop\ III stars were recently supposed to be present in a lensed galaxy showing the faintest Ly$\mathrm{\alpha}$ emission ever observed in the EoR \citep{Vanzella_2020} and in a strong HeII$\lambda 1640$ emitter with extremely blue UV spectral slope at $z = 8.16$ \citep{Wang_2022}, a confirmed detection of this population is still missing.

The combined non-detection of metal emission lines (e.g. [OIII]$\lambda 5007$ and [CIV]$\lambda 1549$) with spectral hardness probes (e.g. HeII$\lambda 1640$ vs H$\alpha$ or H$\beta$) offers a robust method for the identification of first stellar populations inside bright galaxies. Recently, \citealt{Saxena_2020b,Saxena_2020a} drew particular attention to the HeII$\lambda 1640$ emission line, detected in galaxy candidates of the VANDELS\footnote{\url{http://vandels.inaf.it/}} survey at $z \sim 2.3 - 5$, although the candidate sources, either X-ray binaries, weak or obscured active galactic nuclei, or Pop\ III stars, still remain to be clarified. Finally, the relatively short lifetime of very massive stars ($\sim$ few Myr) implies short-lived spectral signatures, making their detection even more challenging.

Theoretical predictions of possible spectroscopic line diagnostics and their combination are already available in the literature \citep{Inoue_2011, Nakajima_Maiolino_2022, Katz_2022, Trussler_2022b}, providing indications on the lifetimes of the signals \citep{Katz_2022} and the level of confusion due to competitive helium-ionizing sources \citep{Nakajima_Maiolino_2022}. However, the rising interest in the above targets stimulates further investigations on the physical properties and statistics of galaxies hosting Pop\ III stars.

According to semi-analytical models \citep{Schneider_2006, Salvadori_2007, deBennassuti_2014, Mebane_2018, Hartwig_2022, Trinca_2022} and numerical simulations \citep{Maio_2009, Bromm_2013, Johnson_2013, Xu_2016_XRB, Liu_Bromm_2020_2, Abe_2021, Sarmento_Scannapieco_2022}, the earliest episodes of Pop\ III star formation occur at $20 \lesssim z \lesssim 30$, in mini-halos with a typical mass of $\sim 10^6 ~ \si{M_\odot}$ \citep{Couchman_1986, Haiman_1996, Tegmark_1997} from the collapse of primordial clouds, primarily caused by $\mathrm{H}_2$ cooling \cite[e.g.][]{Maio_2022}. However, large uncertainties persist on the shape and mass range of their Initial Mass Function (IMF). It is now theoretically established from one-zone models \citep{Omukai_Yoshii_2003, Omukai_2005}, early simulations \citep{Omukai_Nishi_1998, Abel_2002, Bromm_2002, Yoshida_2008} and more recent high-resolution simulations \citep{Hosokawa_2011, Hirano_2014, Hirano_2015_UVrad, Hirano_2015_PPS, Susa_2014, Hosokawa_2016, Sugimura_2020} that inefficient cooling in metal-free gas favours the formation of a top-heavy IMF, with stellar masses ranging from $\sim 10s ~ \si{M_\odot}$ to $\sim 100s ~ \si{M_\odot}$ or even $\sim 1000 ~ \si{M_\odot}$, sometimes with a subdominant tail of low-mass stars \citep{Machida_2008, Clark_2011, Greif_2012, Machida_Doi_2013, Stacy_2016, Susa_2019}. Indirect constraints on the shape of their IMF have been obtained by the metallicity distribution and surface elemental abundances of very metal-poor stars observed in the Galactic halo \citep{Salvadori_2007, deBennassuti_2014, deBennassuti_2017, Hartwig_2018} and nearby dwarf satellites \citep{Salvadori_2008, Rossi_2021, Skuladottir_2021} or both \citep{Graziani_2015, Ishiyama_2016, Magg_2018, Aguado_2023}, favouring a top-heavy IMF.

The transition from a top-heavy Pop\ III IMF to a present-day Pop\ II/I IMF \citep{Salpeter_1955, Kroupa_2002, Chabrier_2003} is believed to be primarily driven by the increase of gas metallicity ($Z_{\rm gas} \gtrsim Z_\mathrm{crit} \sim 10^{-3} - 10^{-4} ~ \si{Z_{\odot}}$, e.g. \citealt{Omukai_2000}, \citealt{Bromm_2001}, \citealt{Maio_2010, Maio_2011}), while a scenario with a dust-driven transition is also plausible, allowing Pop\ II star formation at $Z_\mathrm{crit} \sim 10^{-6} - 10^{-4} ~ \si{Z_{\odot}}$ provided that a critical dust-to-gas mass ratio of $\mathcal{D_\mathrm{crit}} \sim 4.4. \times 10^{-9}$ is reached in star-forming clouds \citep{Schneider_2002, Schneider_2006, Schneider_2012_dustToGas, Schneider_2012_dustSDSSstar, Omukai_2005, Chiaki_2014}. While 3D simulations have confirmed that efficient cooling by metals \citep{Maio_2007} and dust grains \citep{Tsuribe_Omukai_2006, Clark_2008, Dopcke_2011, Dopcke_2013, Safranek-Shrader_2016, Chiaki_2016} promotes fragmentation at small scales, it is hard to follow the subsequent accretion phase for sufficiently long time ($10^4$ - $10^5$ years, \citealt{Chiaki_Yoshida_2022}) to constrain the shape of the stellar IMF. Recently, \citet{Chon_2021} showed that although the number of low-mass stars increases with metallicity when $Z_{\rm gas} \gtrsim 10^{-5}~ \si{Z_{\odot}}$, the stellar IMF converges to a present-day IMF only when $Z_{\rm gas} \gtrsim 10^{-2} ~ \si{Z_{\odot}}$, due to turbulence decay and strong accretion. In addition, the shape of the IMF at $Z_{\rm gas} \sim 10^{-2} - 10^{-1}~ \si{Z_{\odot}}$ is affected by the heating of the Cosmic Background Radiation (CMB) at $z \geq 10$, which suppresses fragmentation reducing the number of low-mass stars \citep{Chon_2022}.

Despite the above findings paint a complex picture on the properties of stellar populations in high-redshift galaxies, numerous studies have described the rate of Pop\ III formation and the global Pop\ III - Pop\ II transition on cosmological scales, by modelling both cosmic metal enrichment and radiative feedback from H-ionizing Ultra-Violet (UV, $> 13.6$~eV) and Lyman-Werner (LW, $11.2-13.6$~eV) photons (see e.g. \citealt{Johnson_2013, Maio_2016, Sarmento_Scannapieco_2022}).

Tracing different atomic metals and cosmic dust through mass and metallicity-dependent stellar yields and accounting for stellar lifetimes and stellar IMFs are necessary ingredients for an accurate description of cosmic metal enrichment \citep{Tornatore_2007_chemicalFeedback, Maio_2010, Graziani_2020}. Primordial non-equilibrium H, He, $\mathrm{H}_2$ and D-based chemistry (\citealt{Galli_Palla_1998}) is crucial as well, in order to determine the physical properties of primordial clouds. Finally, models of galactic winds \citep{Springel_Hernquist_2003} and multi-scale gas mixing \citep{Pan_2013, Sarmento_2016, Sarmento_2017, Sarmento_2018, Sarmento_Scannapieco_2022} are required to follow metals across diffuse environments of the Circum-Galactic/Inter-Galactic Medium (CGM/IGM).
 
Radiative feedback by UV and LW photons is another key process because $\mathrm{H_2}$ photodissociation induced by LW radiation can suppress Pop III star formation in mini-halos \citep{Omukai_Nishi_1999, Machacek_2001, Wolcott-Green_2011, Visbal_2014}, and delay their formation in more massive halos \citep{OShea_Norman_2008, Xu_2013}. In addition, gas heating due to reionization can also prevent star formation in low-mass halos \citep{Okamoto_2008, Sobacchi_Mesinger_2013, Noh_McQuinn_2014, Graziani_2015}.

As the above feedback processes impose severe numerical requirements, different strategies have been adopted depending on the problem at hand. \citealt{Maio_2010}, for example, studied the Pop\ III/II transition  accounting for detailed yields of Pair Instability Supernovae (PISNe), Type II Supernovae (SNeII) and Type Ia Supernovae (SNeIa) from both Pop\ III and Pop\ II stars, and a wide primordial chemistry network. The Renaissance simulations \citep{Xu_2016_latePopIII, Xu_2016_XRB} adopted instead a coupled radiation-hydrodynamics approach by modelling UV and LW radiation from Pop\ III and Pop\ II stars, while adopting a simplified chemical feedback accounting for an averaged metallicity. Despite large differences in the models, hydrodynamical simulations \citep{Pallottini_2014, Jaacks_2019, Skinner_Wise_2020, Sarmento_2018, Sarmento_Scannapieco_2022} indicate that an inhomogeneous metal enrichment drives a smooth statistical transition in redshift as large clumps of pristine gas \citep{Tornatore_2007_PopIII, Maio_2010, Johnson_2013, Xu_2016_latePopIII} can survive in metal-enriched galaxies, allowing Pop\ III star formation even in massive galaxies at the end of the EoR \citep{Liu_Bromm_2020_2, Bennet_Sijacki_2020}.

The present work investigates the statistics of cosmological Pop\ III star formation and the physical properties of galaxies hosting first stars during the EoR. With this aim in mind we adopted the hydrodynamical code \texttt{dustyGadget} \citep{Graziani_2020}, which implements a complex chemical feedback accounting for both atomic metals \citep{Tornatore_2007_PopIII, Tornatore_2007_chemicalFeedback, Maio_2010} and cosmic dust \citep{Graziani_2020}. We also warn the reader that a proper treatment of radiative feedback is currently missing in the \texttt{dustyGadget} model, and the impact of the above limitation will be extensively discussed in the paper. In particular, we analyze Pop III stars present in six hydrodynamical simulations of $50h^{-1} ~ \si{cMpc}$ side length, already introduced in \citet{DiCesare_2022}, by focusing on the statistics of Pop\ III star-forming regions and their galactic environments. Our findings could be useful to guide the search for Pop III stars in upcoming JWST observations.
Two of the Early Release Science (ERS) programs, for example, have observed galaxies well into the EoR and might be able to detect systems potentially hosting Pop\ III stars: the Cosmic Evolution Early Release Science (CEERS, \citealt{Finkelstein_2017, Finkelstein_2022, Finkelstein_2023}) Survey and the Grism Lens-Amplified Survey from Space (GLASS, \citealt{Treu_2017, Treu_2022, Castellano_2022}).

The paper is organised as follows. In Section \ref{sec:simulation}, the \texttt{dustyGadget} code is briefly introduced and the adopted simulation suite described. In Section \ref{sec:results}, our results are presented as follows: Section \ref{sec:results_cosmicSFH}
describes the redshift evolution of the Pop\ III star formation rate density, Section \ref{sec:results_MS} describes how galaxies hosting active Pop\ III stars are placed in the galaxy main sequence, while Sections \ref{sec:results_PopIIIhalos} and \ref{sec:results_LOS} analyze the properties of their star-forming environments. Finally, Section \ref{sec:conclusions} summarizes our conclusions.

\section{Method}
\label{sec:simulation}

\subsection{The \texttt{dustyGadget} code}
\label{sec:simulation_dustyGadget}

The hydrodynamical code \texttt{dustyGadget} \citep{Graziani_2020} is an improved version of \texttt{Gadget-2/3} \citep{Springel_2005_Gadget-2, Springel_2008_Gadget-3} accounting for self-consistent dust production and evolution on top of the chemo-dynamical extensions of the original Smoothed Particle Hydrodynamics (SPH) scheme \citep{Tornatore_2007_chemicalFeedback, Maio_2010, Maio_2011}.

The gas chemical evolution model is inherited from \citet{Tornatore_2007_chemicalFeedback} and follows the metal release from stars with different masses, metallicity and lifetimes, easily implementing alternative metal/dust yields and IMFs. Metals are included in a chemical network accounting for H, He, D and primordial molecules, as described in \citet{Maio_2007}, and allowing stellar-population transition for a given critical metallicity \citep{Tornatore_2007_chemicalFeedback, Maio_2010}.
The gas cooling function \citep{Sutherland_Dopita_1993} consistently reflects the chemical network by accounting for molecular, atomic and fine structure metal transitions of O, $\mathrm{C}^+$, $\mathrm{Si}^+$ and $\mathrm{Fe}^+$ at $T < 10^4 \; \si{K}$ \citep{Maio_2007}. A UV background is also implemented as photo-heating mechanism following \citet{Haardt_Madau_1996}.

Cosmic dust production by Asymptotic Giant Branch (AGB) stars and SNe is described through mass and metallicity-dependent stellar yields, in a way that ensures consistency with the gas-phase metal enrichment. In particular, yields for AGB stars are derived from \citet{Ferrarotti_Gail_2006} and \citet{Zhukovska_2008}, while yields for PISNe and SNeII are adopted from \citet{Schneider_2004} and \citet{Bianchi_Schneider_2007_reverseShock} respectively. Following \citet{Bianchi_Schneider_2007_reverseShock}, we assume that only a fraction of 2 - 20\% of the newly formed SN dust is able to survive the passage of the reverse shock\footnote{The reverse shock destruction efficiency is still very uncertain, as it depends on the nature of the SN explosions, the properties of the grains, and their spatial distribution (see e.g. \citealt{Micelotta_2018} and references therein). The dust yield accounting for reverse shock is also referred to ``effective SN dust yield'', see e.g. \citet{Bocchio_2016_reverseShock}.} and be injected in the Inter-Stellar Medium (ISM). In the current implementation, we follow four classes of dust species: Carbon (C), Silicates ($\mathrm{MgSiO_3}$, $\mathrm{MgSiO_4}$ and $\mathrm{SiO_2}$), Alumina ($\mathrm{Al_2O_3}$) and Iron (Fe). However, it is also possible to implement alternative dust yields (as for the atomic metals) and to include additional grain types. Dust is spread in the ISM together with metals, by using a spline kernel. Finally, the model used for galactic winds comes from \citet{Springel_Hernquist_2003}. Galactic winds are modelled with a constant velocity of $500 ~ \si{km.s^{-1}}$, as suggested by observations of ALPINE normal galaxies \citep{Ginolfi_2020}.

After being produced, dust grains are followed through their evolution across the hot and cold gas phases of the ISM, where they could be destroyed or grow by accretion of metals. At the current stage, \texttt{dustyGadget} does not explicitly follow the evolution of the grain size distribution, but it assumes all the grains are spherical, with a static, average size of $0.1 \; \mu\si{m}$. The interested reader can find more details on the specific processes and their numerical implementation in \citet{Graziani_2020}.

\subsection{Simulation setup}
\label{sec:simulation_dustyGadget_setup}

Our study is carried out on a suite of eight \texttt{dustyGadget} cosmological simulations (hereafter dubbed as U6 - U13)\footnote{Note that data from U9 and U11 has not been included in the present studies as different snapshot dumps have been
used for these cubes with respect to the other ones.} already introduced in \citet{DiCesare_2022}. Each simulated volume has a size of $50h^{-1} \; \si{cMpc}$ and a dark matter/gas particles mass resolution of $3.53 \times 10^7 h^{-1} \; \si{M_\odot}$ / $5.56 \times 10^6 h^{-1}\; \si{M_\odot}$ respectively, for a total number of $2 \times 672^3$ particles. Both volume and resolution have been increased compared to \citet{Graziani_2020} to guarantee a good compromise between an adequate statistics, an acceptable mass resolution and a reasonable computational time per run. To ensure consistency across the simulations, all the volumes share common assumptions for the $\mathrm{\Lambda}$CDM cosmology, consistent with  \citet{Planck_2015} ($\Omega_{\mathrm{m,0}} = 0.3089$, $\Omega_{\mathrm{b,0}} = 0.0486$, $\Omega_{\mathrm{\Lambda},0} = 0.6911$, $h = 0.6774$), and a common feedback setup based on \citet{Graziani_2020}.

We adopt a cold gas phase density threshold for star formation of $n_\mathrm{th} = 132 h^{-2} \; \si{cm^{-3}}$, while the IMF of stellar populations, represented by stellar particles, is assigned according to their metallicity $Z_\star$ \citep{Tornatore_2007_PopIII, Maio_2010, Graziani_2020}, assuming a critical metallicity value of $Z_\mathrm{crit} = 10^{-4} \; \si{Z_\odot}$\footnote{To determine the identity (Pop III or II/I) of the newly-born star particle, the metallicity is calculated like in \citet{Tornatore_2007_PopIII}, by smoothing on the SPH kernel normalized in a sub-region of the SPH volume (0.2 of the SPH length). In addition, we assume $Z_\mathrm{\odot} = 0.02$ \citep{Anders_Grevesse_1989}.}. In particular, for the present study we have made the following choices for the stellar IMF:
\begin{enumerate}
    \item for $Z_\star < Z_\mathrm{crit}$ (Pop\ III stars), a Salpeter IMF \citep{Salpeter_1955} in the mass range [100, 500] \si{M_\odot} is adopted. Mass-dependent yields describe the metal pollution from stars in the PISN range [140, 260] \si{M_\odot} \citep{Heger_Woosley_2002};
    \item for $Z_\star \geq Z_\mathrm{crit}$ (Pop\ II/I stars), a standard Salpeter IMF \citep{Salpeter_1955} in the mass range [0.1, 100] \si{M_\odot} is adopted. Mass and metallicity-dependent yields describe the metal pollution from long-lived, low-intermediate mass stars \citep{vanDenHoek_Groenewegen_1997}, high mass stars (> 8 \si{M_\odot}), dying as core-collapse SNe \citep{Woosley_Weaver_1995}, and SNeIa \citep{Thielemann_2003}.
\end{enumerate}

When stars enrich the surrounding gas, we follow the evolution of six metals (C, O, Mg, S, Si and Fe). We note here that the current model for Pop III stars only considers enrichment from PISNe. Potential signatures of PISNe have been found on the surface elemental abundances of metal-poor stars in our Galactic neighbourhood \citep{Aoki_2014, Salvadori_2019, Aguado_2023}, and in the broad line region gas of one of the most distant quasars at $z = 7.54$ \citep{Yoshii_2022}. However, the properties of carbon-enhanced extremely metal-poor stars are best matched by the nucleosynthetic output by Pop III faint supernovae \citep{Ishigaki_2014, deBennassuti_2014, deBennassuti_2017, Fraser_2017, Magg_2022, Aguado_2023a}, implying Pop III stellar progenitors with masses $\sim 10 - 40 \, M_\odot$. In future studies, we will explore alternative shapes of the Pop III IMF and how these affect the Pop III star formation history (see also the discussion at the end of Appendix~\ref{sec:AppSFRD}).
We also note that Pop\ III stars outside the PISN mass range and Pop\ II/I stars with masses $\geq 40 \; \si{M_\odot}$ are assumed to directly collapse into black holes and therefore they do not contribute to the chemical enrichment. Black holes are not followed explicitly.

We safely assume that Pop\ II stellar populations have a long tail of low-mass stars that can survive across cosmic times, while Pop\ III stars, due to their high mass, die after a relatively short time. The average lifetime of a Pop\ III star is: 
\begin{equation}
    \overline{\tau} = \frac{\int_{m_\mathrm{low}}^{m_\mathrm{up}} \tau(m) \phi(m) \dd{m}}{\int_{m_\mathrm{low}}^{m_\mathrm{up}} \phi(m) \dd{m}} \simeq 3 \; \si{Myr},
    \label{eq:lifetime}
\end{equation}
\noindent
where $\phi(m)$ is the adopted IMF, $m_\mathrm{low}$ and $m_\mathrm{up}$ are the lower and upper stellar mass limits, and $\tau (m)$ is the mass-dependent main sequence lifetime taken from \citet{Schaerer_2002}. Hence, a system is classified as a ``Pop\ III halo'' when it hosts Pop\ III stars younger than 3 Myr.

All the simulations are carried out from $z \simeq 100$ down to $ z \simeq 4$, while this work will focus on the redshift range $z \geq 6.5$; indeed, we expect a significant drop in Pop\ III star formation in the post-reionization epoch due to the combined effect of cosmic metal enrichment and UV/LW radiative feedback. Finally, the identification of Dark Matter (DM) halos and their substructures is performed in post-processing with the \texttt{AMIGA} halo finder (AHF, \citealt{Knollmann_Knebe_2009}).

Note that a good statistics is required for our investigation in order to reproduce the cosmic star formation history of all stellar populations (see Section \ref{sec:results_cosmicSFH}), to ensure reliability in the predicted galaxy scaling relations \citep{DiCesare_2022}, as well as to guarantee a number of moderately resolved galaxy candidates with Pop\ III star forming regions. The ability to disentangle star-forming environments, in particular, is necessary to explore the internal properties of some of these candidates (Section \ref{sec:results_PopIIIhalos}). Moreover, it should be noted that having more independent cubes allows us to investigate the impact of cosmic variance on our results, which is important especially at high redshifts ($z \gtrsim 13$, see Section \ref{sec:results_cosmicSFH}), and to access a wider statistics and variety of star-forming galaxies. This is a key point when looking for candidates hosting episodic Pop\ III star formation.

\section{Results}
\label{sec:results}

Here we discuss our simulation results. In Section \ref{sec:results_cosmicSFH} the cosmic star formation rate introduced in \citet{DiCesare_2022} is revisited to highlight the contributions of Pop\ III and Pop\ II stars during EoR. The statistics of main sequence galaxies hosting Pop\ III star-forming regions is discussed in Section \ref{sec:results_MS}, while their properties are investigated in Sections \ref{sec:results_PopIIIhalos} and \ref{sec:results_LOS}.

\subsection{Pop\ III cosmic star formation history}
\label{sec:results_cosmicSFH}

\begin{figure*}
\centering
\includegraphics[angle=0, width=.85\textwidth, clip]{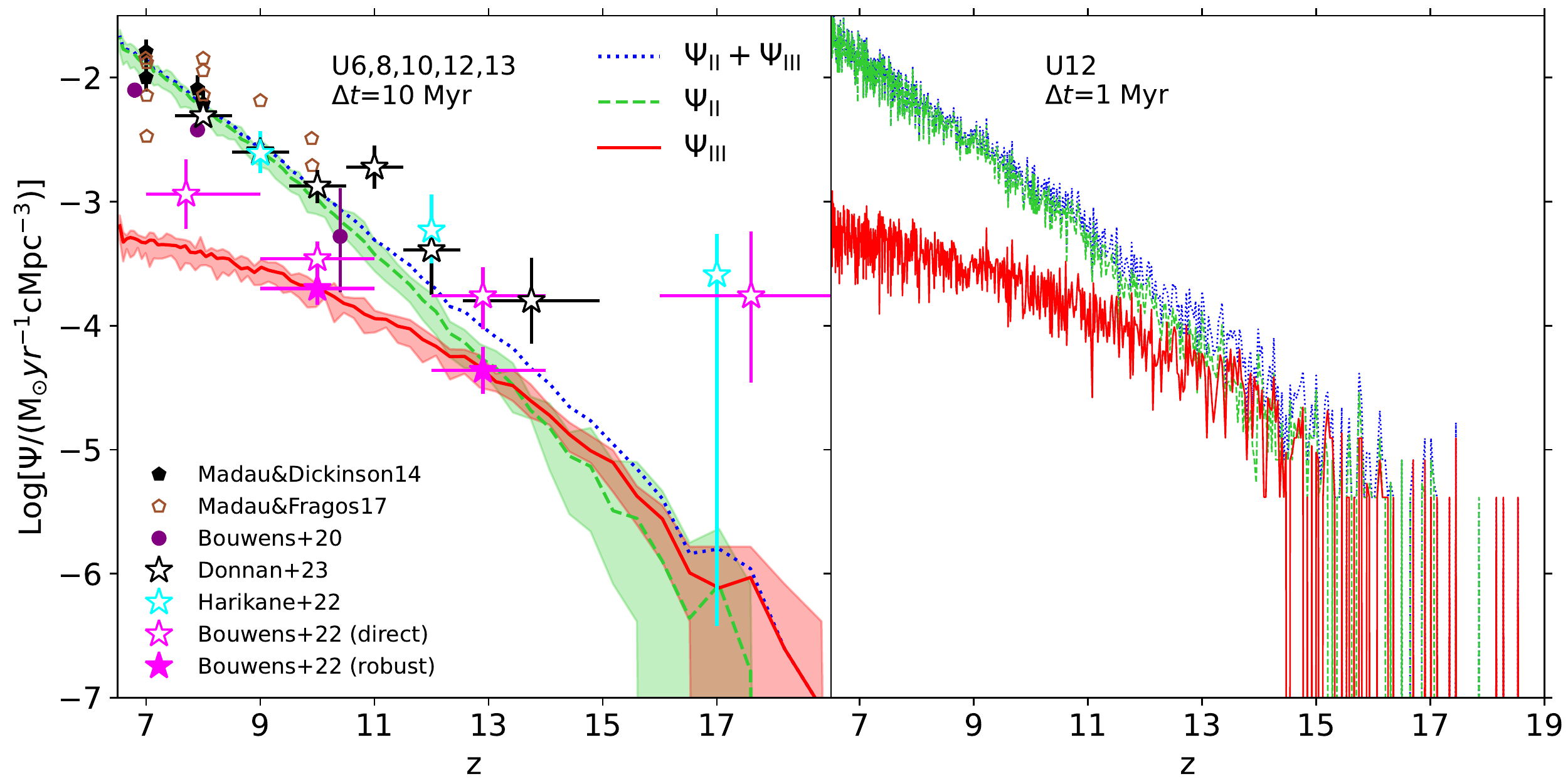}
\caption{The comoving SFRD of Pop\ III ($\Psi_\mathrm{III}$), Pop\ II ($\Psi_\mathrm{II}$) stars and their sum ($\Psi_\mathrm{II}$+$\Psi_\mathrm{III}$) as a function of redshift, in the range $6.5 \leq z \leq 19$. In the \textbf{left panel}, the SFRD is computed with a timestep $\Delta t = 10 ~ \si{Myr}$, averaged through the simulated cubes U6, U8, U10, U12, U13. Particularly, the mean value of $\Psi$ for Pop\ III (Pop\ II) stars is represented by a \textit{red solid} (\textit{green dashed}) line, while the \textit{red} (\textit{green}) shaded area indicate the spread between the minimum and maximum values across the cubes; the mean value of the total SFRD is also shown by a \textit{blue dotted} line. Results for the total SFRD are compared with observed points taken from \citet{Madau_Dickinson_2014} (\textit{black, filled pentagons}), \citet{Madau_Fragos_2017} (\textit{sienna, empty pentagons}), \citet{Bouwens_2020} (\textit{purple, filled circles}), \citet{Donnan_23} (\textit{black, empty stars}), \citet{Harikane_2023} (\textit{cyan, empty stars}) and \citet{Bouwens_2023} (\textit{hotpink, filled/empty stars}). As for the latter, the filled stars (``direct'') are direct estimates from \citet{Bouwens_2023} fiducial reductions of JWST data in the SMACS0723 \citep{Pontoppidan_2022}, GLASS Parallel \citep{Treu_2017, Treu_2022, Castellano_2022} and CEERS \citep{Finkelstein_2017, Finkelstein_2022, Finkelstein_2023} fields, while the empty stars (``robust'') are derived from literature galaxy samples at $z \gtrsim 8$ in the same field, for which the estimated cumulative probability that the candidates lie at $z > 5.5$ exceeds 99\%. The \textbf{right panel} shows the evolution of $\Psi$ in the simulated cube U12, computed with a timestep $\Delta t = 1 ~ \si{Myr}$, using the same color legend adopted in the left panel.}
\label{fig:SFRD_pop_av+obs+U12}
\end{figure*}

This section investigates the relative contributions of Pop\ III and Pop\ II stars to the cosmic Star Formation Rate Density (SFRD, $\Psi$) predicted by our simulation suite. \citet{DiCesare_2022} have already shown that the SFRD predicted by \texttt{dustyGadget} simulations is in good agreement with data at $z \geq 4$ and first estimates from the JWST Early Science data. Here we perform a step further by investigating the formation rates of Pop\ III ($\Psi_{\rm III}$) and Pop\ II ($\Psi_{\rm II}$) stars, in order to understand early star formation on cosmic scales.

Differently from \citet{DiCesare_2022}, the total Star Formation Rate (SFR) is derived by discretizing the time in temporal steps $\Delta t$, starting at $z \sim 20$, and by computing the stellar mass $\Delta M_\star$ formed in each interval $\Delta t$ by looking at the birth time of all the stellar particles, without considering their association with DM halos. Hence, the SFRD of the $i$-th cube at time $t$ is given by:
\begin{equation}
    \Psi_i(t) = \frac{1}{V_\mathrm{c}}\frac{\Delta M_\star(t, t - \Delta t)}{\Delta t},
\end{equation}
where $\Delta M_\star(t, t - \Delta t)$ is computed by summing over the masses of all the stellar particles with an age between $t$ and $t - \Delta t$, and $V_\mathrm{c} = (50 h^{-1} ~ \si{cMpc})^3$ is the comoving volume. Considering that the star formation process occurring on the smaller timesteps of the hydrodynamical simulation is stochastic, lower $\Delta t$ values result in more significant fluctuations of $\Psi$. Moreover, choosing an appropriate value of $\Delta t$ is key to capture the formation of massive, short-lived Pop\ III stars. For the above reasons, the redshift evolution of $\Psi$ is investigated adopting two values of $\Delta t$:  $\Delta t = 1$ Myr (i.e. smaller than $\overline{\tau} = 3$~Myr, the average lifetime of Pop\ III stars), and $\Delta t = 10$~Myr.

The right panel of Figure \ref{fig:SFRD_pop_av+obs+U12} shows $\Psi = \Psi_{\rm II} + \Psi_{\rm III}$ (dotted, blue line), $\Psi_{\rm II}$ (dashed, green line) and $\Psi_{\rm III}$ (solid, red line) as a function of redshift, as predicted in U12\footnote{U12 is chosen as a prime example to illustrate the behaviour of $\Psi$ in a single simulated cube. U12 is also the universe with the earliest onset of star formation.} assuming $\Delta t = 1 ~ \si{Myr}$. The values of $\Psi$ adopting $\Delta t = 10 ~ \si{Myr}$ and averaging across universes are shown instead in the left panel; for an easier comparison with the right panel, the same line styles are adopted, while the spread across different simulated volumes is indicated as red and green shaded areas (for $\Psi_\mathrm{III}$ and $\Psi_\mathrm{II}$, respectively). As a reference, a selected set of recent observations \citep{Bouwens_2020} and data-constrained estimates \citep{Madau_Dickinson_2014, Madau_Fragos_2017} are reported. Recent estimates from JWST Early Release Science \citep{Donnan_23, Harikane_2023, Bouwens_2023} are also included in the plots, but see \citet{DiCesare_2022} for a more extended comparison with observations.

In the redshift range $15 < z \leq 19$ the first episodes of star formation mix Pop\ III and Pop\ II stars in U12 (right panel), while at $z < 15$ the SFRD starts to increase for both populations. The ratio $\Psi_\mathrm{II} / \Psi_\mathrm{III}$ rapidly increases at $z \leq 13$, reaching a value of $\Psi_\mathrm{II} / \Psi_\mathrm{III} \sim 30$ at $z = 6.5$. $\Psi_{\rm III}$ exhibits significant fluctuations (up to one order of magnitude) among consecutive $\Delta t$. Towards the end of EoR ($z \lesssim 7$), when Pop\ III star formation is disfavoured by increased metal pollution, the average $\Psi_{\rm III}$ flattens around $\Psi_{\rm III} \sim 10^{-3.2} \, {\rm M}_\odot \, {\rm yr}^{-1} \, {\rm cMpc}^{-3}$, with oscillations $\sim 10\%$.

By adopting a $\Delta t = 10 ~ \si{Myr}$ and by averaging across simulations (left panel), the fluctuations of $\Psi$ are significantly suppressed in both populations, and a global statistical trend emerges. The first episodes of star formation occur, across universes, in the redshift range $16 \lesssim z \lesssim 19$ and are dominated by Pop\ III stars. At these redshifts, a non-negligible spread around the average $\Psi_\mathrm{III}$ can be appreciated\footnote{It is well known that the results on Pop III star formation are strongly influenced by resolution and by the inability to resolve low-mass galaxies. Due to the low mass resolution, few regions might be able to satisfy the star formation criterion at high redshifts ($z \gtrsim 15$), resulting in a later offset of star formation and in a large scatter of both Pop\ III and Pop\ II SFRD (also see the right panel of Figure~\ref{fig:SFRD_pop_av+obs+U12}). A detailed study of the effect of resolution on the SFRD at high redshifts is out of the scopes of the present work.}, and Pop\ III star formation is confirmed to be highly stochastic in all the simulated volumes across cosmic time, due to the inhomogeneous nature of cosmic metal enrichment. The Pop\ II contribution to the SFRD increases with time, finally overcoming the Pop\ III contribution around $z \simeq 13$. The values of $\Psi_{\rm II}$ found in different simulated cubes at these redshifts are well convergent and become gradually closer to the mean value (dashed, green line). While being subdominant, $\Psi_{\rm III}$ continues to persist with a value of $\Psi_{\rm III} \lesssim 10^{-3} \; \si{M_\odot.yr^{-1}.cMpc^{-3}}$ in all cubes during the EoR, and by $z = 6.5$ the evolution $\Psi_{\rm III}$ has significantly flattened, without showing a decline.

Pop\ III star formation is mainly suppressed by cosmic metal enrichment and, locally, by thermal feedback from supernova explosions. It is important to remind the reader that, at present, \texttt{dustyGadget} does not account for a proper Radiative Transfer (RT) in the photo-dissociating LW and UV ionizing radiation and for a proper metal mixing below our gas mass resolution, which has been shown to enhance Pop III star formation by a factor 2-3 \citep{Sarmento_2016, Sarmento_2017, Sarmento_2018, Sarmento_Scannapieco_2022}. Sub-grid turbulent metal diffusion can also be important for the metallicity distribution functions in dwarf galaxies \citep{Escala_2018}, by driving individual gas and star particles towards the average metallicity, although \cite{Su_2017} showed that this has little impact on general star formation and ISM properties. \cite{Jeon_2017} further demonstrate that the effect of diffusion is weaker at lower metallicity and relatively unimportant for Pop III star formation.

Note that the applied homogeneous UV background does not have feedback on cosmic star formation at $z > 6$, i.e. in the epoch considered in the present work. In addition, locally strong LW radiation field created by stars in the same galaxy would suppress late Pop III star formation in some primordial-gas pockets by the $\mathrm{H}_2$ photodissociation (\citealt{Omukai_Nishi_1999, Johnson_2013, Maio_2016}; also refer to the discussion in Appendix~\ref{sec:AppSFRD}. These limitations certainly affect the late-time behaviour of $\Psi_{\rm III} (z)$, i.e. our results should be interpreted as upper limits of $\Psi_{\rm III}$ at these redshifts.

\subsubsection{Comparison with large-scale models}
\label{sec:results_cosmicSFH_comparisons}

\begin{figure}
    \centering
    \includegraphics[width=\linewidth]{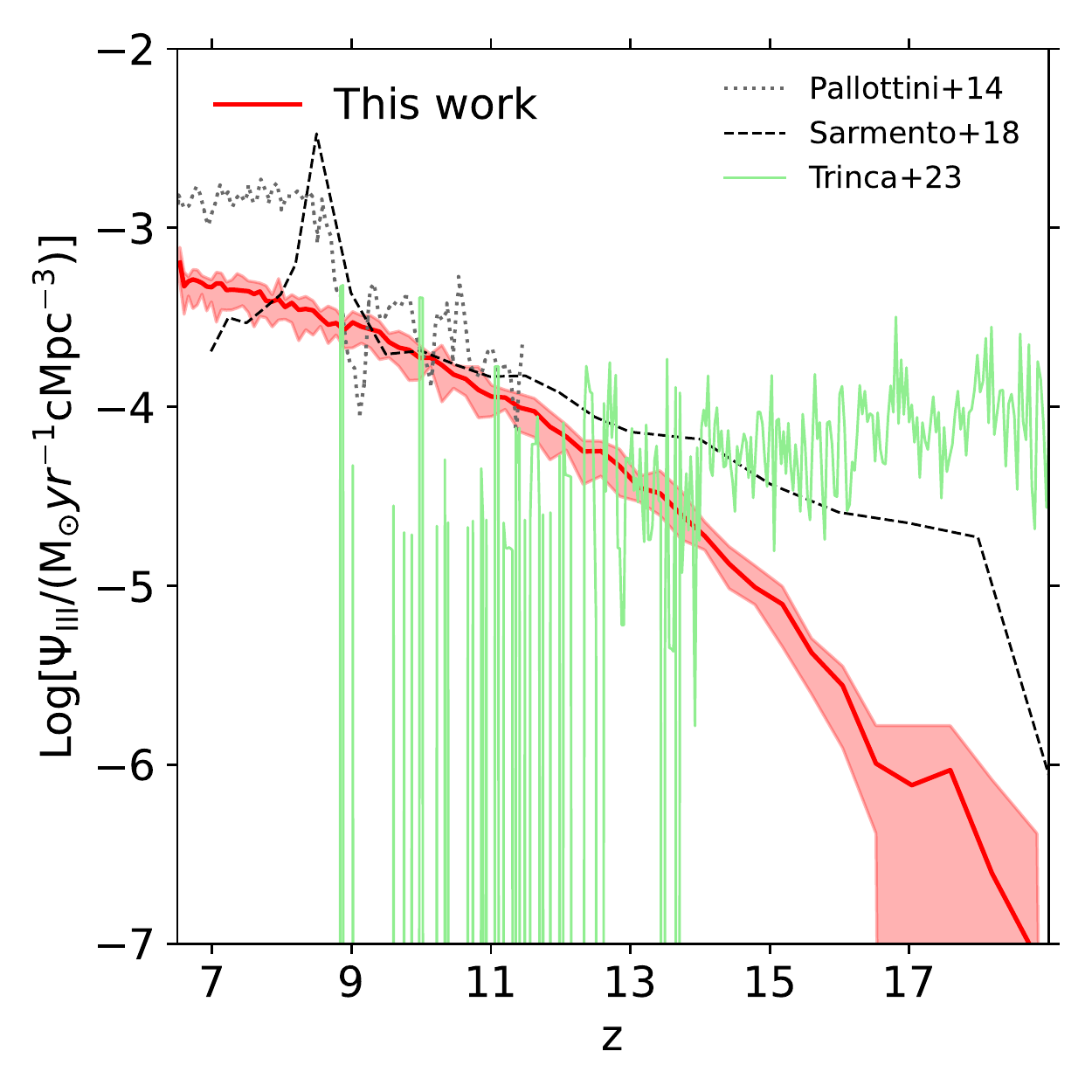}
    \caption{Average Pop\ III SFRD $\Psi_\mathrm{III}$ (computed as in the left panel of Figure \ref{fig:SFRD_pop_av+obs+U12}) compared with the results of other large-scale (box sizes $\geq 10h^{-1} ~ \si{cMpc}$) models and simulations, i.e. \citet{Pallottini_2014} (RAMSES simulation, \textit{grey, dotted lines}), \citet{Sarmento_2018} (RAMSES simulation, \textit{black, dashed line}) and Trinca et al. in prep (CAT semi-analytic model, \textit{light-green, solid line}). See Table \ref{tab:PopIII_models} for a compilation of the models used as a comparison here and in Figure \ref{fig:SFRD_pop_av+modSmallScale}, with their main features.}
    \label{fig:SFRD_pop_av+modLargeScale}
\end{figure}

As Pop\ III star formation is still not constrained by observations, interesting clues on model improvements and result reliability can come from the comparison with independent theoretical predictions available in the literature.

A comparison of $\Psi_{\rm III}(z)$ predicted by models and simulations with a box size $L \geq 10h^{-1}$~cMpc is shown in Figure \ref{fig:SFRD_pop_av+modLargeScale}. First note that the values of the SFRD at $z \gtrsim 13$ are strongly dependent on the adopted mass/scale resolution: up to three orders of magnitude difference is found between $\Psi_{\rm III}(z\geq 13)$ predicted by \texttt{dustyGadget} and the one resulting from higher-resolution simulations, or from semi-analytical models capable of resolving star formation in mini-halos.
For example, the higher SFRD found by \citet{Sarmento_2018} (black, dashed line), is likely due to a better resolved star formation in mini-halos\footnote{Their mass resolution is sufficient to resolve DM halos with masses of $\sim 10^7 ~ \si{M_\odot}$. For further details, we refer to the original paper.}. Below $z \sim 13$, when star formation is less sporadic and more sustained in a larger population of galaxies hosted in Ly$\alpha$-cooling halos, a reasonable agreement between our predictions and the trends of \citet{Sarmento_2018} and \citet{Pallottini_2014} is found\footnote{The sudden increase of $\Psi_{\rm III}(z\sim 9)$, common to both AMR methods, is probably of numerical origin, and difficult to justify in terms of feedback effects.}.

From the comparison with \texttt{CAT} (light-green, solid line), it is evident that all numerical methods capable to resolve the internal hydrodynamics of star-forming halos and adopting an inhomogeneous metal enrichment predict late Pop\ III star formation, extending towards the end of the EoR. In fact, a sharper, and probably overly anticipated transition to Pop\ II stars is common to semi-analytical models not implementing inhomogeneous metal enrichment (e.g. \citealt{Graziani_2017}, Trinca et al. in prep.); other models taking into account this effect appear to be more in line with the results of hydrodynamical simulations (see e.g. the agreement of \citealt{Visbal_2020} with simulations on smaller boxes in Appendix~\ref{sec:AppSFRD}). Finally, we note that a proper multi-frequency treatment of RT (as the one adopted in RT codes, e.g. \citealt{Eide_2018, Eide_2020}), still not present in the above cosmological models, should flatten $\Psi_{\rm III}(z \leq 7)$ and progressively drop its values by suppressing star formation in mini-halos or small Ly$\alpha$-cooling halos, at least at the end of reionization ($z\sim 5.5$, see e.g. \citealt{Eide_2020}), when a uniform LW and UV background is established on the large scale of the IGM.

Despite all the aforementioned differences, and the resulting spread among model predictions, a common feature to most of the models presented in Figure \ref{fig:SFRD_pop_av+modLargeScale} is that they predict a late phase of Pop\ III star formation, extending down to $z \sim 7$, albeit with values of $\Psi_{\rm III}(z)$ largely subdominant with respect to the dominant Pop\ II contribution (see Figure \ref{fig:SFRD_pop_av+obs+U12} for a reference). Interestingly, a considerable fraction of these late Pop\ III star-forming regions is found in globally enriched, Pop\ II-forming galaxies as a result of a predicted persistence of pristine regions in globally, but inhomogeneously enriched environments (also see Section \ref{sec:results_MS}). A comparison with models and simulations on smaller scales is presented in Appendix~\ref{sec:AppSFRD}.

\subsection{Pop\ III star formation in main sequence galaxies}
\label{sec:results_MS}

This section investigates the properties of Pop\ III-forming, Main Sequence (MS) galaxies predicted by \texttt{dustyGadget}, selected to have stellar masses\footnote{This corresponds to halos with $\gtrsim 20$ stellar particles; indeed, we do not rely on systems described by a small number of stellar particles, where our results are confused by numeric noise and the halo finder algorithm is not able to properly recognise the halos found in specific mass ranges.} $\mathrm{Log} (M_\star/ \rm M_{\odot}) \geq 7.5$, evolving in the EoR, at $6.7 \leq z \leq 8.1$; these systems may be observable by JWST and ALMA. While an extended discussion of the MS can be found in \citet{DiCesare_2022}, hereafter we focus on potential MS galaxies hosting Pop\ III stars, to study their star-formation properties and ISM environments.

\begin{figure*}
    \centering
    \includegraphics[angle=0, width=\textwidth, trim={0 0 0 0},clip]{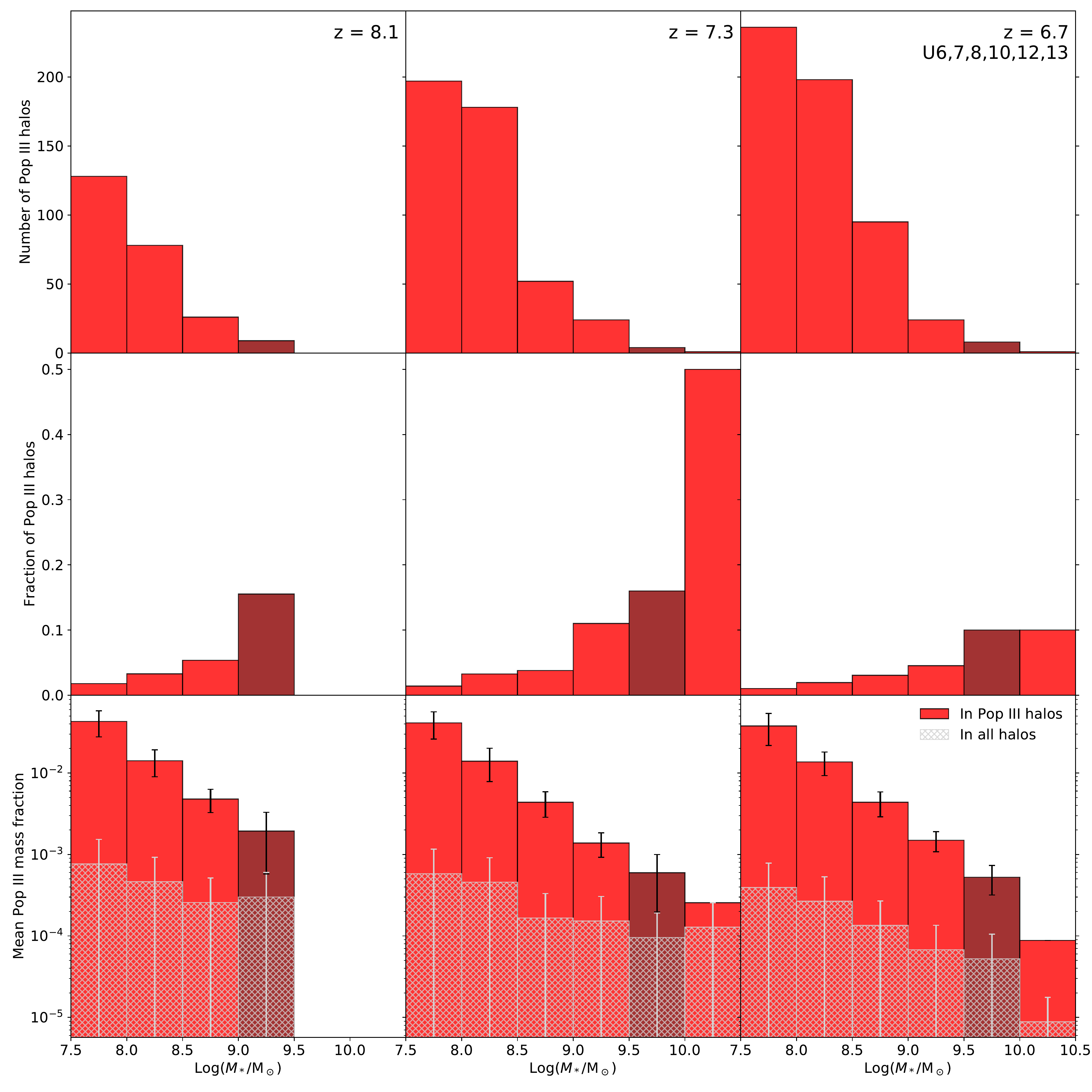}
    \caption{\textbf{Top panels:} number of Pop\ III halos in six bins of stellar mass $M_\star$ (with a spacing of 0.5~dex in the range $7.5 \leq \mathrm{Log} M_\star / \si{M_\odot} < 10.5$). \textbf{Middle panels:} number fraction of Pop\ III halos in the same $M_\star$ bins. \textbf{Bottom panels}: mean Pop\ III mass fraction in Pop\ III halos (\textit{red bins}) and in all halos (\textit{white, hatched bins}) in the same $M_\star$ bins; also shown is the standard deviation of each bin. Results are shown for the simulated cubes U6, U7, U8, U10, U12, U13 at $z = 8.1$ (\textbf{left panel}), $z = 7.3$ (\textbf{middle panel}) and $z = 6.7$ (\textbf{right panel}). For each redshift, the bins containing the Pop\ III halos selected for a further study of their internal properties in Section \ref{sec:results_PopIIIhalos} (lines in bold in Table \ref{tab:PopIIIhalos_globalProperties}) are highlighted in \textit{dark-red}.}
    \label{fig:histograms_popIII}
\end{figure*}

 Figure~\ref{fig:histograms_popIII} shows the distribution of Pop\ III-forming galaxies as a function of their stellar mass $M_\star$ for the simulated volumes U6, U7, U8, U10, U12 and U13 at redshifts $z = 8.1, \, 7.3$ and 6.7. All the objects included in these plots are mixed systems, hosting both Pop\ III and Pop\ II stellar populations, while Pop\ III-only systems - with no Pop\ II - are all in the range of stellar masses where the simulated galaxies are no longer well resolved ($\mathrm{Log} (M_\star / \si{M_\odot}) < 7.5$).

The histograms in the top and middle panels show the number and number fraction of Pop\ III hosts in different stellar mass bins. The distributions indicate that, while the absolute number of Pop\ III halos decreases with stellar mass, their relative fraction increases in rare massive galaxies, up to a value $\sim 50 \%$ for halos with $M_\star \geq 10^{10} ~ \si{M_\odot}$ at $z = 7.3$. Therefore, according to our model and in light of providing clues to upcoming observations, high-mass, Pop\ III - Pop\ II mixed halos at these redshifts are promising Pop\ III hosts. In the bottom panels of Figure~\ref{fig:histograms_popIII} we also show the average Pop\ III mass fraction found in Pop\ III hosts, which is decreasing with stellar mass\footnote{Notably, the semi-analytic study of \cite{Riaz_2022} also shows a similar trend, although at higher redshifts ($z \gtrsim 12$). See for example their Figures 2 and 3.}. Hence, our results suggest that the optimal observational strategy to identify Pop\ III stars would be to find the best compromise between the brightest Pop\ III hosts and their highest Pop\ III mass fraction.

\begin{figure*}
    \centering
    \includegraphics[angle=0,width=\textwidth, trim={0 0 0 0},clip]{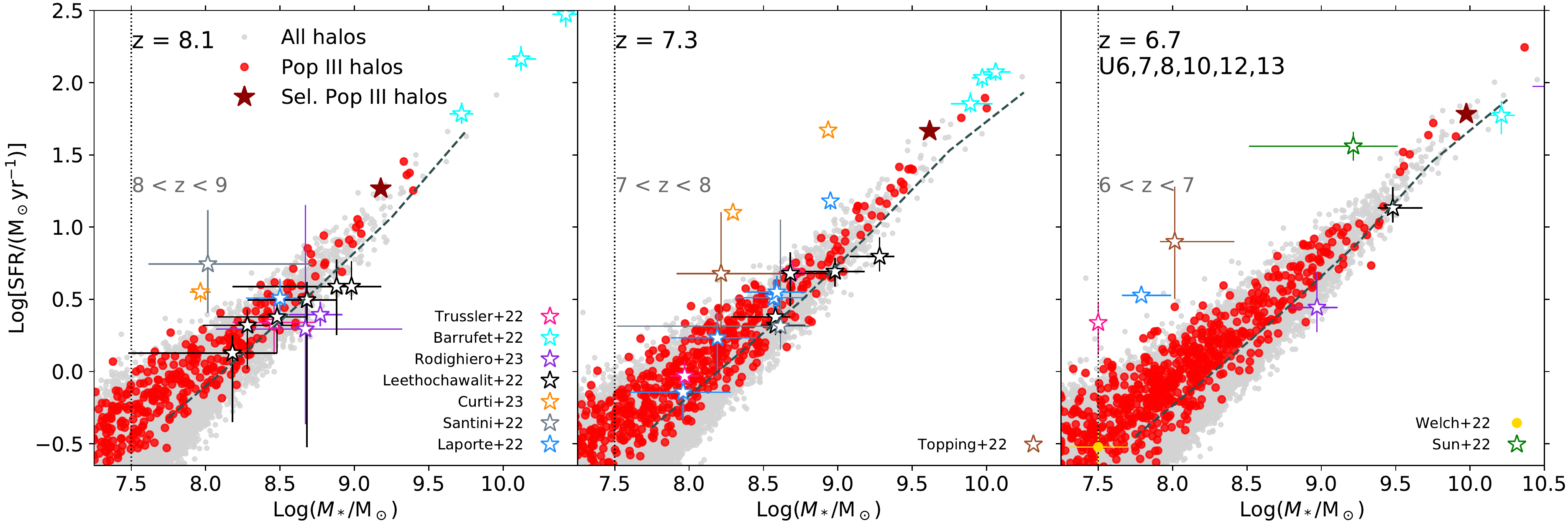}
    \caption{MS of star formation for the general halo population (\textit{light-grey, filled dots}) and for the Pop\ III halos (\textit{red, filled dots}) in the simulated cubes U6, U7, U8, U10, U12, U13 at $z = 8.1$ (\textbf{left panel}), $z = 7.3$ (\textbf{middle panel}) and $z = 6.7$ (\textbf{right panel}). The \textit{thin-dotted, vertical, black lines} mark our resolution limit ($\mathrm{Log} M_\star / \si{M_\odot} \geq 7.5$). The \textit{black, dashed lines} show the mean values of the SFRs for the general population in six bins of stellar mass $M_\mathrm{*}$ (with a spacing of 0.5~dex in the range $7.5 \leq \mathrm{Log} M_\star / \si{M_\odot} < 10.5$). The Pop\ III halos selected for a further spatially resolved study of their internal properties in Section \ref{sec:results_PopIIIhalos} (lines in bold in Table \ref{tab:PopIIIhalos_globalProperties}) are highlighted by \textit{dark-red stars}. Empty stars with \textit{pink, cyan, purple, black, orange, grey, cerulean, brown and dark-green} borders show comparisons with the recent results obtained from JWST early data release \citep{Trussler_2022, Barrufet_2023, Rodighiero_2023, Leethochawalit_2022, Curti_2023, Santini_2022, Laporte_2022, Topping_2022_JWST, Sun_2022}. The Sunrise Arc \citep{Welch_2022} is also shown as a \textit{yellow circle} in the right panel. We apply the proper conversion factors \citep{Madau_Dickinson_2014} to $M_\star$ and SFRs that were originally computed with an IMF different from the Salpeter one.} 
    \label{fig:MS_popIII}
\end{figure*}

As second step of our analysis, in Figure \ref{fig:MS_popIII} we show the MS of star-forming galaxies found in the same simulated cubes at $z = 8.1$, $z = 7.3$ and $z = 6.7$. Each point in the SFR vs $M_\star$ plane corresponds to a single central galaxy of DM halos identified by the halo finder. Pop\ III-forming objects are coloured as red dots to show their position on the MS compared to the total population of star-forming galaxies, represented by light-grey dots. Note that data in Figure~\ref{fig:histograms_popIII} and \ref{fig:MS_popIII} share the same bins of $M_\star$, and the stellar-mass resolution threshold $\mathrm{Log} (M_\star / \si{M_\odot}) = 7.5$ is indicated as thin-dotted, vertical, black lines.

\begin{table*}
    \centering
    \caption{List of the Pop\ III halos chosen for a study of their internal properties (Section \ref{sec:results_PopIIIhalos}) at redshifts $z = 8.1$, $z = 7.3$ and $z = 6.7$, selected among the most massive candidates of our simulated volumes ($M_\star \gtrsim 10^9 ~ \si{M_\odot}$). For each candidate, the halo ID, stellar mass $\mathrm{Log} (M_\star / \si{M_\odot})$, Pop\ III mass to total stellar mass ratio $\mathrm{Log} (M_\mathrm{III} / M_\star)$, SFR [\si{M_\odot.yr^{-1}}], dust mass $\mathrm{Log} (M_\mathrm{dust} / \si{M_\odot})$, gas mass $\mathrm{Log} (M_\mathrm{gas} / \si{M_\odot})$ and dark matter mass $\mathrm{Log} (M_\mathrm{DM} / \si{M_\odot})$ are displayed in the various columns. The halos selected for a further spatially resolved study in Section \ref{sec:results_PopIIIhalos} are marked in \textit{bold}.}
    \begin{tabular}{ccccccc}
        \hline
        Halo ID & $\mathrm{Log} (M_\star / \si{M_\odot})$ & $\mathrm{Log} (M_\mathrm{III} / M_\star)$ & SFR [\si{M_\odot.yr^{-1}}] & $\mathrm{Log} (M_\mathrm{dust} / \si{M_\odot})$ & $\mathrm{Log} (M_\mathrm{gas} / \si{M_\odot})$ & $\mathrm{Log} (M_\mathrm{DM} / \si{M_\odot})$ \\
        \hline
        \multicolumn{7}{c}{$z = 8.1$}\\
        \hline
        U12H3 & 9.40 & -3.08 & 18 & 7.15 & 10.42 & 11.20 \\
        U8H2 & 9.37 & -3.06 & 24 & 6.88 & 10.65 & 11.39 \\
        U12H1 & 9.35 & -3.04 & 23 & 6.80 & 10.63 & 11.37 \\
        U8H1 & 9.33 & -2.93 & 28 & 6.76 & 10.68 & 11.44 \\
        \textbf{U7H4} & \textbf{9.18} & \textbf{-2.26} & \textbf{18} & \textbf{6.58} & \textbf{10.55} & \textbf{11.30} \\
        U8H5 & 9.02 & -2.61 & 11 & 6.46 & 10.45 & 11.20 \\
        U10H2 & 9.01 & -2.70 & 10 & 5.95 & 10.43 & 11.18 \\
        U10H5 & 8.99 & -2.68 & 8 & 6.51 & 10.22 & 10.99 \\
        U6H5 & 8.97 & -2.65 & 8 & 6.25 & 10.37 & 11.12 \\
        \hline
        \multicolumn{7}{c}{$z = 7.3$}\\
        \hline
        U6H0 & 10.00 & -3.59 & 67 & 7.83 & 10.85 & 11.66 \\
        U8H0 & 9.99 & -3.68 & 78 & 7.78 & 11.08 & 11.85 \\
        U8H1 & 9.83 & -3.52 & 57 & 7.46 & 10.96 & 11.70 \\
        \textbf{U7H2} & \textbf{9.62} & \textbf{-2.91} & \textbf{46} & \textbf{7.16} & \textbf{10.81} & \textbf{11.56} \\
        U12H2 & 9.50 & -3.19 & 25 & 7.05 & 10.81 & 11.57 \\
        U8H4 & 9.49 & -3.18 & 25 & 7.21 & 10.66 & 11.41 \\
        U6H2 & 9.44 & -3.13 & 18 & 7.14 & 10.63 & 11.40 \\
        U7H3 & 9.44 & -2.83 & 20 & 6.76 & 10.80 & 11.55 \\
        U10H2 & 9.41 & -2.80 & 26 & 6.86 & 10.63 & 11.40 \\
        U13H2 & 9.38 & -3.07 & 17 & 6.87 & 10.59 & 11.35 \\
        U13H0 & 9.35 & -2.69 & 19 & 6.92 & 10.73 & 11.46 \\
        U10H3 & 9.33 & -3.02 & 15 & 6.87 & 10.62 & 11.35 \\
        U7H95 & 9.00 & -2.38 & 7 & 6.10 & 10.18 & 10.59 \\
        \hline
        \multicolumn{7}{c}{$z = 6.7$}\\
        \hline
        U8H0 & 10.37 & -4.05 & 176 & 8.27 & 11.25 & 12.04 \\
        \textbf{U8H1} & \textbf{9.97} & \textbf{-3.66} & \textbf{61} & \textbf{7.61} & \textbf{11.19} & \textbf{11.94} \\
        U7H6 & 9.90 & -3.50 & 42 & 7.75 & 10.85 & 11.61 \\
        U13H0 & 9.75 & -3.44 & 53 & 7.29 & 10.97 & 11.73 \\
        U8H54 & 8.94 & -2.63 & 6 & 5.97 & 10.42 & 11.20 \\
        \hline
    \end{tabular}
    \label{tab:PopIIIhalos_globalProperties}
\end{table*}

The scatter plots suggest that Pop\ III halos are systematically shifted towards higher SFRs\footnote{The location of Pop\ III halos with respect to the mass-metallicity relation (e.g. \citealt{Langeroodi_2022, Curti_2023}) would further demonstrate the conditions of late Pop III star formation. We defer to Cataldi et al. in prep. for a study of metallicity in \texttt{dustyGadget} galaxies through an appropriate estimator, taking into account the star-forming regions of the galaxies while excluding contaminations from the surrounding CGM.} (they mostly lie above the mean values of the general sample, i.e. the black dashed lines), and that they tend to follow a tighter SFR vs $M_\star$ relation. This interesting segregation may point out to specific conditions favoring late Pop\ III star formation, such as recent episodes of pristine gas accretion, which decrease the metallicity and induce star formation, and/or particularly quiet and smooth star formation histories, as opposed to bursty and stochastic, which lead to lower level of metal enrichment and mixing in galaxies with comparable stellar masses and redshifts.

By analyzing the properties of a sub-sample of massive objects, we will address the above points again in the next section, finding plausible observational targets. In Figure \ref{fig:MS_popIII}, we also report recent JWST observations\footnote{Note that the data from \citet{Leethochawalit_2022} have been updated with respect to \citet{DiCesare_2022} to match the published version of the paper.}  \citep{Trussler_2022, Barrufet_2023, Rodighiero_2023, Leethochawalit_2022, Curti_2023, Santini_2022, Laporte_2022, Topping_2022_JWST, Sun_2022}, which clearly indicate that a relevant part of our candidate sample with masses $8.0 \leq \mathrm{Log} (M_\star/\rm M_{\odot}) < 9.0$ and $\mathrm{Log} (M_\star/\rm M_{\odot}) \geq 9.0$ can be observed with JWST at these redshifts, where additional galaxy samples have also been targeted by ALMA (see for example the recent REBELS survey, \citealt{Bouwens_2022_REBELS}).

Since a large fraction of alive Pop\ III stellar mass is predicted by \texttt{dustyGadget} to be in poorly-resolved systems (e.g. $\gtrsim 60 \%$ of Pop\ III stellar mass at $z = 6.7$ is in halos with $\mathrm{Log} (M_\star/ \si{M_\odot}) \lesssim 7.5$, and this percentage grows up to $\sim 90\%$ at $z = 10$), an increased mass resolution and a more refined feedback prescription are certainly required to investigate Pop\ III star formation at high redshifts ($z \gtrsim 10$), and to quantify the possibility of a persisting mode of Pop\ III star formation in unresolved mini-halos during the EoR ($z \lesssim 10$). \citet{Liu_Bromm_2020_2}, for example, find that the contribution of mini-halos to the overall Pop\ III SFRD drops to a few percent at $z \lesssim 13$, due to LW feedback; thereafter, it decreases exponentially down to $z = 0$, where $\Psi_{\rm III} \lesssim 10^{-5} ~ {\rm M_\odot}\,{\rm yr}^{-1}\,{\rm cMpc}^{-3}$. In agreement with our large-scale predictions, the latest Pop\ III star formation episodes occur in low-metallicity pockets of more massive halos (with a virial mass $\gtrsim 10^9 ~ \si{M_\odot}$), due to inefficient metal mixing. 

Finally, we need to remind that the above findings may be affected by mass resolution and that the detection of a stellar population inside a high-$z$ object is subject to a series of uncertainties. Indeed, in smaller and coarser simulated halos, the gas component is traced by few particles that are quickly affected by chemical enrichment as soon as the first stars release their metals. In these halos, the simulation is not capable to provide a sufficiently accurate description of the spatial distribution of gas and metals; hence, the ISM properties are averaged out, similarly to what is commonly assumed in semi-analytic models. In the next section we scrutinize the properties of a selected sample of galaxies providing clues of plausible Pop\ III candidate hosts.

\subsection{Pop\ III halo properties during the EoR}
\label{sec:results_PopIIIhalos}
In this section we discuss the properties of 27 Pop\ III hosts\footnote{As a reference, note that a total of 241, 456 and 562 resolved ($\mathrm{Log} (M_\star/ \rm M_{\odot}) \geq 7.5$) Pop III halos have been found in the simulated cubes U6, U7, U8, U10, U12 and U13 at $z = 8.1, 7.3$ and 6.7 respectively.} extracted from the \texttt{dustyGadget} catalog at redshifts $z = 8.1$, $z = 7.3$ and $z = 6.7$. The galaxies are selected among the most massive and best resolved halos in the simulated cubes U6, U7, U8, U10, U12 and U13, in order to characterize their internal properties and surrounding environments. Table \ref{tab:PopIIIhalos_globalProperties} lists the selected candidates, together with their global properties: total stellar mass ($M_\star$), fraction of stellar mass in Pop\ III stars ($M_\mathrm{III} / M_\star$), instantaneous SFR at target redshift, total dust and gas mass ($M_\mathrm{dust}$, $M_\mathrm{gas}$), and mass of the DM halo ($M_\mathrm{DM}$). 

From an observational perspective, several massive star-forming objects of this kind have already been identified in the EoR. 322 high-redshift galaxy candidates at $5.5 < z < 8.5$ are available from the Hubble and Spitzer Space Telescope  observational campaign RELICS (Reionization Lensing Cluster Survey, \citealt{Salmon_2020}), among which the so-called ``Sunrise Arc'' at $z = 6.2$ has recently become famous for the detection of a supposed individual, persistently magnified star, Earendel\footnote{\citet{Schauer_2022} investigated the possibility that Earendel is a Pop\ III star. Starting from the simulation of \citet{Liu_Bromm_2020_2}, they estimated that the probability that Earendel is a Pop\ III is non-negligible throughout the inferred mass range ($M_\star > 50 ~ \si{M_\odot}$, \citealt{Welch_2022}), although the most likely outcome remains a Pop II origin.} (\citealt{Welch_2022}). Recently, the ALMA REBELS large program (Reionization Era Bright Emission Line Survey, \citealt{Bouwens_2022_REBELS}) has targeted 40 among the most UV-bright galaxies at $6.5 < z < 9.5$. With the advent of JWST, more candidates have already been identified from the ERS programs, especially from CEERS (Cosmic Evolution Early Release Science, \citealt{Finkelstein_2017, Finkelstein_2022, Finkelstein_2023}) and GLASS (Grism Lens-Amplified Survey from Space, \citealt{Treu_2017, Treu_2022, Castellano_2022}). 

\begin{figure}
    \centering
    \includegraphics[width=\linewidth]{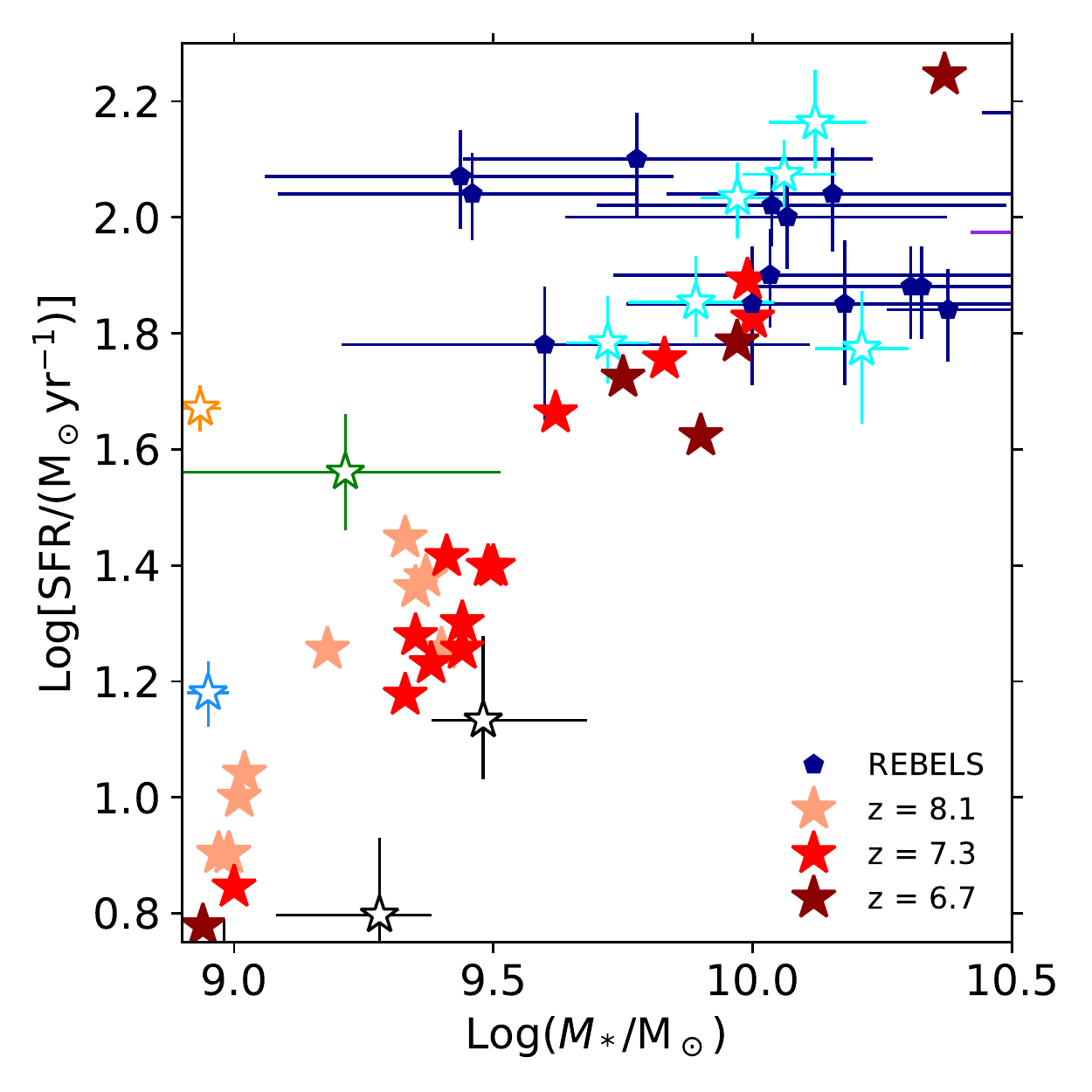}
    \caption{Comparison between the Pop\ III halos listed in Table \ref{tab:PopIIIhalos_globalProperties} and observations in the SFR vs stellar mass ($M_\star$) plane. The simulated systems are shown as filled stars at $z = 8.1$ (\textit{light-red}), $z = 7.3$ (\textit{red}) and $z = 6.7$ (\textit{dark-red}). The observed data points are the same as in Figure \ref{fig:MS_popIII} (we have adopted the same colour/symbol), with the addition of galaxies from the REBELS survey (\citealt{Topping_2022_REBELS}, \textit{blue pentagons}).}
    \label{fig:MS_popIIIselected}
\end{figure}

Figure \ref{fig:MS_popIIIselected} places our tabulated targets in the context of available observations. The simulated systems are indicated as filled stars, with colors corresponding to three different redshifts, while JWST observations are shown with the same color/symbol coding adopted in Figure \ref{fig:MS_popIII}; REBELS galaxies \citep{Topping_2022_REBELS} are also shown as blue pentagons. At the present stage of our investigation, we do not make any claim on the presence or observability of Pop\ III stars in any of the reported galaxies. However, the above figure provides a global indication that rare MS galaxies with $\mathrm{Log} (M_\star/ \rm M_{\odot}) \gtrsim 9.0$ and $\mathrm{SFR} \gtrsim 10 ~ \si{M_\odot.yr^{-1}}$ ($\sim 2-5\%$ of our resolved - $\mathrm{Log} (M_\star/\rm M_{\odot}) \geq 7.5$ - galaxy population at these redshifts), currently accessible to many observational facilities including ALMA and JWST, could potentially host Pop\ III stars, providing important indications on the properties of star-forming regions and stellar populations in the first Gyr of galaxy formation \citep{Tacchella_2022}. Furthermore, by exploring massive systems in statistically equivalent simulation volumes, we can enlarge the sample of galaxies at the high-mass end of the MS, and investigate their individual star formation histories (see also Figure~\ref{fig:SFHHalos_dt1.0Myr+inset}).

\subsubsection{From the cosmic web to star-forming regions}
\label{sec:results_PopIIIhalosCosmicWebToSFRegions}

Hereafter, we focus on the Pop\ II/III galactic properties of halos U7H4 at $z = 8.1$, U7H2 at $z = 7.3$ and U8H1 at $z = 6.7$ (indicated in bold in Table \ref{tab:PopIIIhalos_globalProperties} and with dark-red histogram bins/stars in Figure~\ref{fig:histograms_popIII}/\ref{fig:MS_popIII}).

\begin{figure*}
    \centering
    \includegraphics[angle=0,width=\textwidth, trim={0 0 0 0},clip]{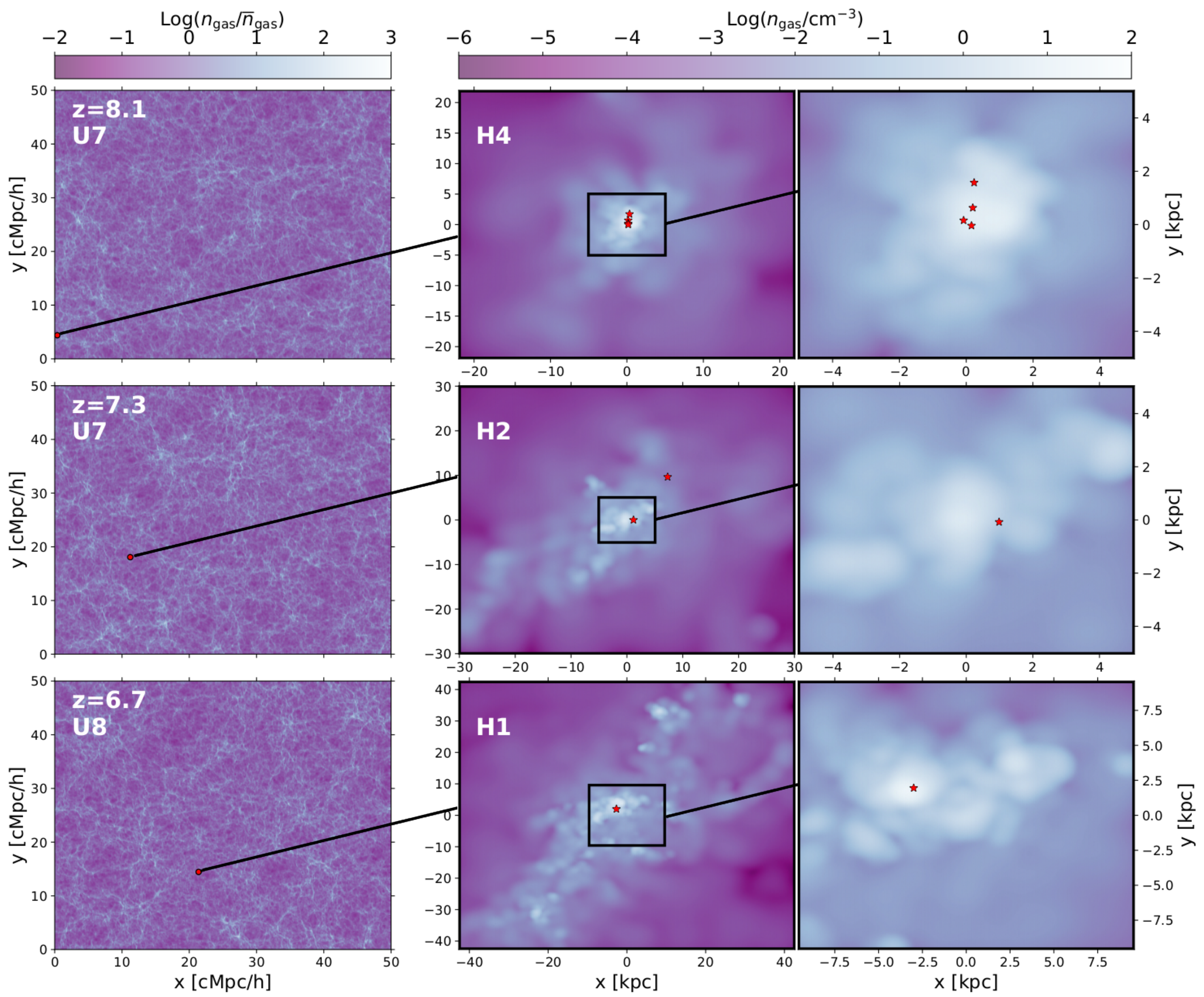}
    \caption{\textbf{Left panels:} maps of the gas number density $n_\mathrm{gas}$ along a slice cut perpendicular to the $z$-axis of the comoving cosmological cubes normalised to the the average number density $\overline{n}_\mathrm{gas}$ of the cube, passing through the Pop\ III halos of interest, i.e. U7H4 at $z = 8.1$ (\textbf{top panels}), U7H2 at $z = 7.3$ (\textbf{middle panels}) and U8H1 at $z = 6.7$ (\textbf{bottom panels}). The positions of the halos on these planes are highlighted in \textit{red}. \textbf{Middle panels:} maps of $n_\mathrm{gas}$ along a slice cut perpendicular to the $z$-axis of each halo on a scale of the order of two times its virial radius (see text for details), passing through one of its Pop\ III stellar particles. \textbf{Right panels:} \textit{zoom-in} view of the middle panels. The projected position of Pop\ III stellar particles on these planes are highlighted in \textit{red}.}
    \label{fig:sliceCuts_plotMatrix}
\end{figure*}
\begin{figure*}
    \centering
    \includegraphics[width=\linewidth]{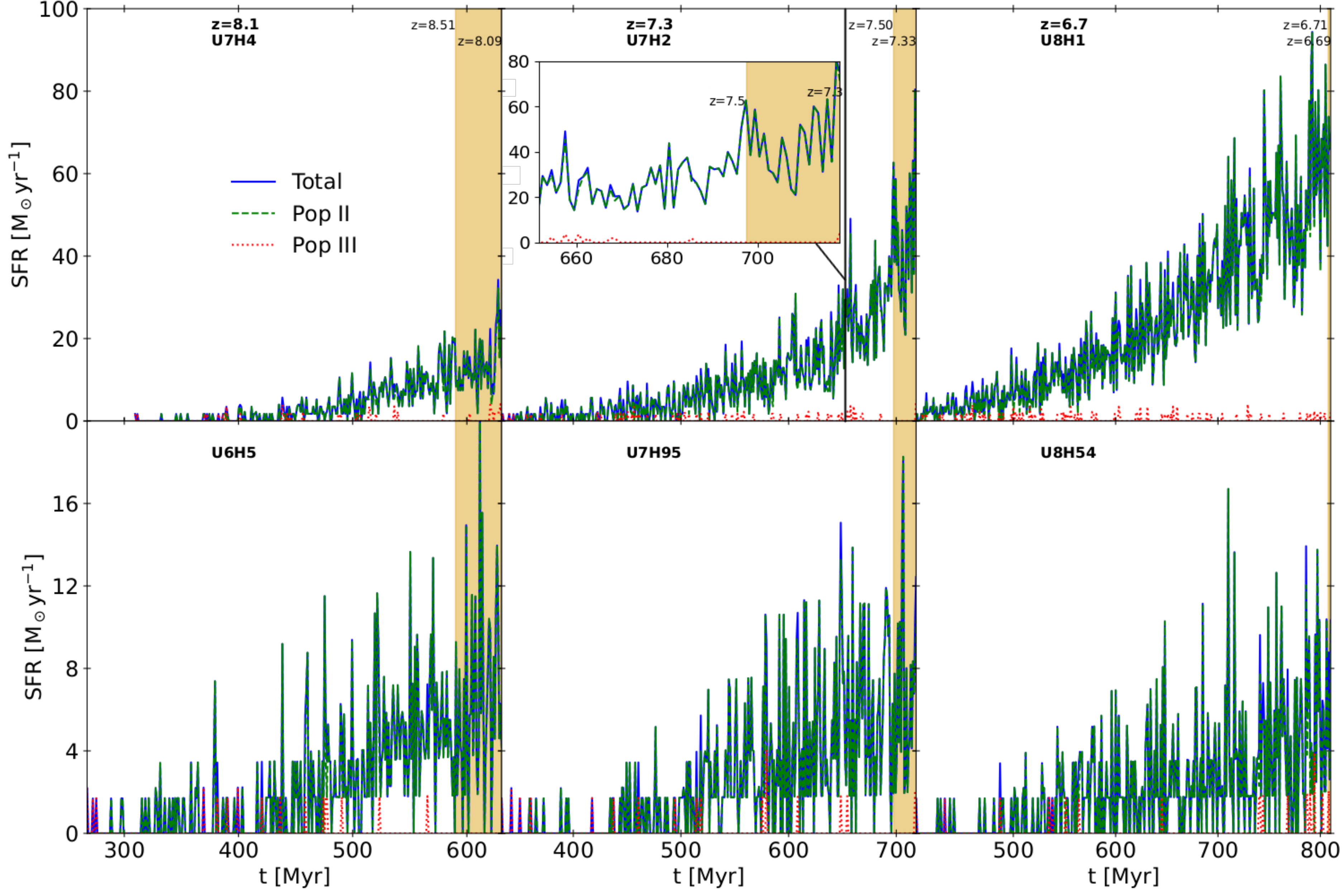}
    \caption{SFR evolution in some of our candidate halos at $z = 8.1$ (\textbf{left column}) $z = 7.3$ (\textbf{middle column}) and $z = 6.7$ (\textbf{right column}), computed in timesteps of 1~Myr along the time range from the birth of the first stellar population in the halo to the time of the current snapshot. \textbf{Top row:} halos selected for a further study of their internal properties in Section \ref{sec:results_PopIIIhalos}, i.e. U7H4, U7H2 and U8H1. \textbf{Bottom row:} least massive halos at each of the three redshifts from Table \ref{tab:PopIIIhalos_globalProperties}, i.e. U6H5, U7H95 and U8H54. The total SFR is represented by \textit{solid, blue} lines, the Pop\ II SFR by \textit{dashed, green} lines and the Pop\ III SFR by \textit{dotted, red} lines. The \textit{shaded, golden regions} illustrate the range between the current and the previous snapshots of the simulations. The \textbf{inset} in the top, middle panel also shows a zoom-in view of the evolution in U7H2 at later times. Note the different range of values in the y-axis in the top e bottom rows.}
    \label{fig:SFHHalos_dt1.0Myr+inset}
\end{figure*}

Figure~\ref{fig:sliceCuts_plotMatrix} shows maps of the gas number density surrounding/inside the above halos on different spatial scales. Gas particles are projected onto cartesian grids with 512 cells/side at increasing spatial resolution (from left to right columns), and each particle contribution is weighted with the SPH kernel adopted in the simulations. In particular, in the first column we show the halo position on the global scale of the simulated volume ($50 h^{-1} ~ \si{cMpc}$), visualising a slice cut passing through the halo and perpendicular to the $z$-axis. The adopted grids have a spatial resolution of $97.7h^{-1} ~ \si{ckpc}$. In the selected planes, the gas distribution is shown in adimensional over-density units, computed as the gas number density $n_\mathrm{gas}$ normalised to the cosmic average value $\overline{n}_\mathrm{gas}$.

The second and third columns progressively zoom into each halo structure, to highlight the position of Pop\ III stellar particles (red stars) on top of the $n_\mathrm{gas}$ distribution. The grids adopted in the middle panel resolve the virial radius of each DM halo ($r_\mathrm{vir} = 22.0 , \, 30.1, \, 42.7 ~ \si{kpc}$ for U7H4, top, U7H2, middle, and U8H1, bottom panel, respectively) and have a corresponding spatial resolution of 85.9~pc, 118~pc and 167~pc. Finally, in the right column the adopted grids resolve squared areas of 26.0, 36.3 and 49.9~kpc/side around the galaxy centre (i.e. with resolutions of 50.8, 70.9 and 97.5 pc), from top to bottom, intercepting at least one Pop\ III star-forming region. Regions containing the halos at different scales are connected through progressively zoomed-in views by a line to guide the eye.

These halos have been chosen because they provide prototypical examples of interesting classes of Pop\ III halos. More specifically, U8H1 at $z = 6.7$ contains a Pop\ III stellar population close to its center, in a crowded region, with other more metal-enriched stellar populations. U7H2 hosts Pop\ III star-forming regions in different galactic environments: near the centre and in an isolated, pristine gas cloud placed at the outskirt of the halo. Finally, U7H4 hosts the highest number of Pop\ III stellar populations (i.e. four Pop\ III stellar particles), providing an example of the highest Pop\ III mass fraction possibly expected at these epochs. Note that all the selected halos are found in over-dense regions close to knots of the baryonic cosmic web, corresponding to values of $n_\mathrm{gas}/\overline{n}_\mathrm{gas} \gtrsim 1000$. As a reference, the lowest/highest overdensities in the cubes correspond to $\sim 0.01/2200$ ($z = 8.1$), $\sim 0.02/1900$ ($z = 7.3$) and $\sim 0.004/4500$ ($z = 6.7$).

In the mid panels, the complex structure of various gas clumps hosted in the DM halos can be appreciated by following the gradients of $n_\mathrm{gas}$ (see colour palette). A comparison between the three selected halos, which have DM mass\footnote{As a reference, a Milky Way-like DM halo at $z=0$ has a $M_{\rm DM}\sim 1.5 - 1.9 \times 10^{12} ~ \si{M_\odot}$, see Appendix A of \citet{Graziani_2017} and references therein.} ${\rm Log} M_{\rm DM}/M_\odot = 11.30 - 11.96$, and redshift $6.7 - 8.1$, shows an increasing level of gas clumpiness with cosmic time, corresponding to different stages of galaxy assembly. This, in turn, corresponds to an increasing number of potentially star-forming regions, to an increased level of metal enrichment, and to a reduced availability of Pop\ III star-forming sites. The largest number of Pop\ III star-forming regions are found towards the centre of the system at $z = 8.1$, indicating that in U7H4 only a mild level of metal pollution is present in the central region, where the higher densities favor star formation. 

As halos assemble, and progressively more gas clumps starts to be present, the number of Pop\ III star-forming regions decreases at the centre and sporadic episodes are allowed in the outskirts of the halos, as shown by the case of U7H2 at $z = 7.3$. Interestingly, the case of U8H1 (bottom row) shows that even towards the end of cosmic reionization, at $z = 6.7$, a limited number of very low-metallicity star-forming regions can still survive close to the galaxy center, allowing sporadic formation of Pop\ III stars. Due to the mass resolution of the simulations, we cannot zoom-in further onto star-forming regions, and this limits our possibility to resolve gas structures on scales smaller than $\sim 50 ~\si{pc}$\footnote{The impact of mass resolution on these scales has also been discussed by \citet{Glatzle_2022}. In this study, a random cloud remodelling has been implemented to capture spatial scales smaller than the typical size of giant molecular clouds (see in particular the discussion of their Figure 7).}.

\subsubsection{Pop\ II/III star formation histories}
\label{sec:results_PopIIIhalosSFH}

Figure~\ref{fig:SFHHalos_dt1.0Myr+inset} shows the Star Formation History (SFH) of 6 halos selected from Table~\ref{tab:PopIIIhalos_globalProperties}. The SFR is computed in timesteps of 1~Myr, starting from the 
birth of the first stellar population in the halo, down to the time corresponding to the simulation snapshot where
the halo was identified. The top row shows the SFHs of the halos marked in bold in Table~\ref{tab:PopIIIhalos_globalProperties} (also see Figure~\ref{fig:sliceCuts_plotMatrix}), while in the bottom row the SFHs of galaxies at same redshift but with the lowest $M_\star$ (i.e. U6H5, U7H95 and U8H54) are reported.

In each panel, the total SFH is shown as a solid blue line, while Pop\ II and Pop\ III contributions are represented by dashed, dark-green and dotted, red lines respectively. The shaded, golden regions identify the interval between the current and the previous snapshots of the simulations, where the Pop\ III populations presented in Table~\ref{tab:PopIIIhalos_globalProperties} and in Figure~\ref{fig:sliceCuts_plotMatrix} (for the three halos in the upper row) are formed. In the case of U7H2, a panel inset provides a close-up of the SFH in which its Pop\ III stars are formed (see also Table~\ref{tab:PopIII_localEnvironments} for more details on the properties of Pop\ III stars).

The SFHs of the three halos represented in the top panels show that massive galaxies (i.e. with ($\mathrm{Log} M_\star/\si{M_{\odot}}) \gtrsim 9$) living in the EoR can experience different SFHs depending on their environments and assembly status. After an almost flat SFH with $\mathrm{SFR} \lesssim 5 ~ \si{M_\odot.yr^{-1}}$ (characterized by few sporadic peaks at $\sim 10 ~ \si{M_\odot.yr^{-1}}$), the SFH of U7H2 and U8H1 rapidly increases by a factor 8 in the last 200~Myr. In the case of U7H4 at $z\sim 8.1$, the galaxy is seen at the beginning of the rapid ascent described above. As already noticed in Figure~\ref{fig:sliceCuts_plotMatrix}, while the total stellar mass becomes rapidly dominated by Pop\ II stars, the three galaxies continue to host sporadic, yet subdominant episodes of Pop\ III star formation at all epochs (see dotted red lines in top panels). 

Coeval, lower-mass galaxies (Log$(M_\star / \si{M_\odot}) \lesssim 9$), chosen as comparison cases (shown in the bottom panels of Figure~\ref{fig:SFHHalos_dt1.0Myr+inset}), exhibit instead flatter, highly stochastic and inefficient SFHs, which hardly ever reach values as high as $20 ~ \si{M_\odot.yr^{-1}}$. Despite these smaller systems experience an earlier onset of star formation compared to their more massive counterparts, their inefficient SFR persists down to the final redshift. This suggests that the birth environment - rather than the redshift at which star formation starts - determines the growth rate of these systems. Pop\ III star-forming episodes appear with lower frequency in these inefficiently star-forming objects, although randomly and at all epochs during the EoR.

\subsubsection{Pop\ II/III star-forming environments}
\label{sec:results_PopIIIhalosSFE}

Hereafter, we will discuss some global halo/galactic properties of selected Pop\ III hosts on physical scales of $\sim 10 ~ \si{kpc}$.

\begin{figure*}
    \centering
    \includegraphics[angle=0,width=\textwidth, trim={0 0 0 0},clip]{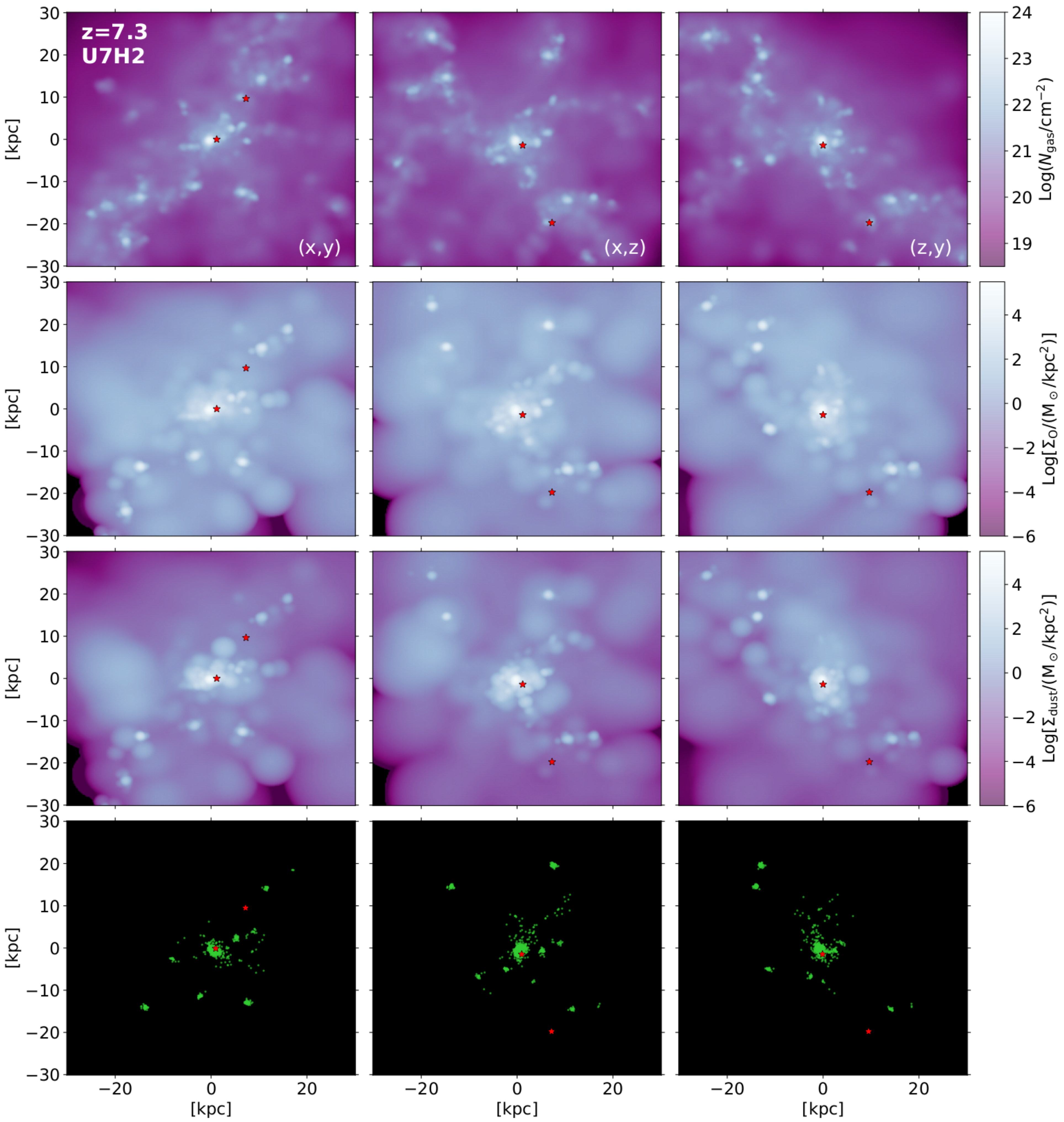}
    \caption{\textbf{Top three panels:} maps of the gas ($N_\mathrm{gas}$), oxygen ($\Sigma_\mathrm{O}$) and dust ($\Sigma_\mathrm{dust}$) column/surface densities in halo U7H2 at $z = 7.3$, along a line-of-sight parallel 
    to the $z$, $y$ and $x$-axis (\textbf{left}, \textbf{middle} and \textbf{right panels}, respectively). 
    The \textbf{bottom panels} show collapsed projections of the Pop\ II stellar particles in the halo along the 
    same lines-of-sight, in \textit{green}. Each map has a side of the order of two times the virial radius of the halo (refer to the beginning of Section \ref{sec:results_PopIIIhalos} for details). The projected position of Pop\ III stellar particles on these planes are highlighted by \textit{red stars}.}     \label{fig:snap020_U7H2_340.0_collapsedProjection_plotMatrix}
\end{figure*}

The very irregular and clumpy nature of halo U7H2 at $z = 7.3$ can be easily appreciated in Figure~\ref{fig:snap020_U7H2_340.0_collapsedProjection_plotMatrix}, where we show maps of its gas column density ($N_\mathrm{gas}$, first row) and oxygen/dust surface densities ($\Sigma_\mathrm{\rm O}$/$\Sigma_\mathrm{dust}$, second/third rows respectively). Each of the columns show the distributions as seen through lines-of-sight parallel to a reference frame axis, $z$ (left), $y$ (middle), $x$ (right). In the bottom row we show the projected positions of stellar particles (indicated as green dots) in order to appreciate their spatial correlation with the previous quantities. In all the panels, the positions of Pop\ III stellar particles are represented as red stars.

Although the highest gas densities are found in a $\sim 1 ~ \si{kpc}$ region close to the centre of mass of the system, corresponding to the position of the central galaxy, many gas clumps are distributed along filaments extending out to distances of $\geq 10 ~ \si{kpc}$ from the centre, possibly tracing the assembly process of the central object. Interestingly, by comparing their positions with the stellar sources (bottom panel), it is evident that many of them are probably small satellite dwarfs with active star formation. As a consequence, the entire halo environment is significantly enriched by atomic metals - as traced by the oxygen abundance -, and dust grains, although their spatial distributions do not always correlate with each other, probably due to dust destruction and sputtering processes operating in the shocked and high-temperature CGM. 

\begin{figure*}
    \centering
    \includegraphics[angle=0,width=0.9\textwidth, trim={0.1cm 0.1cm 0 0},clip]{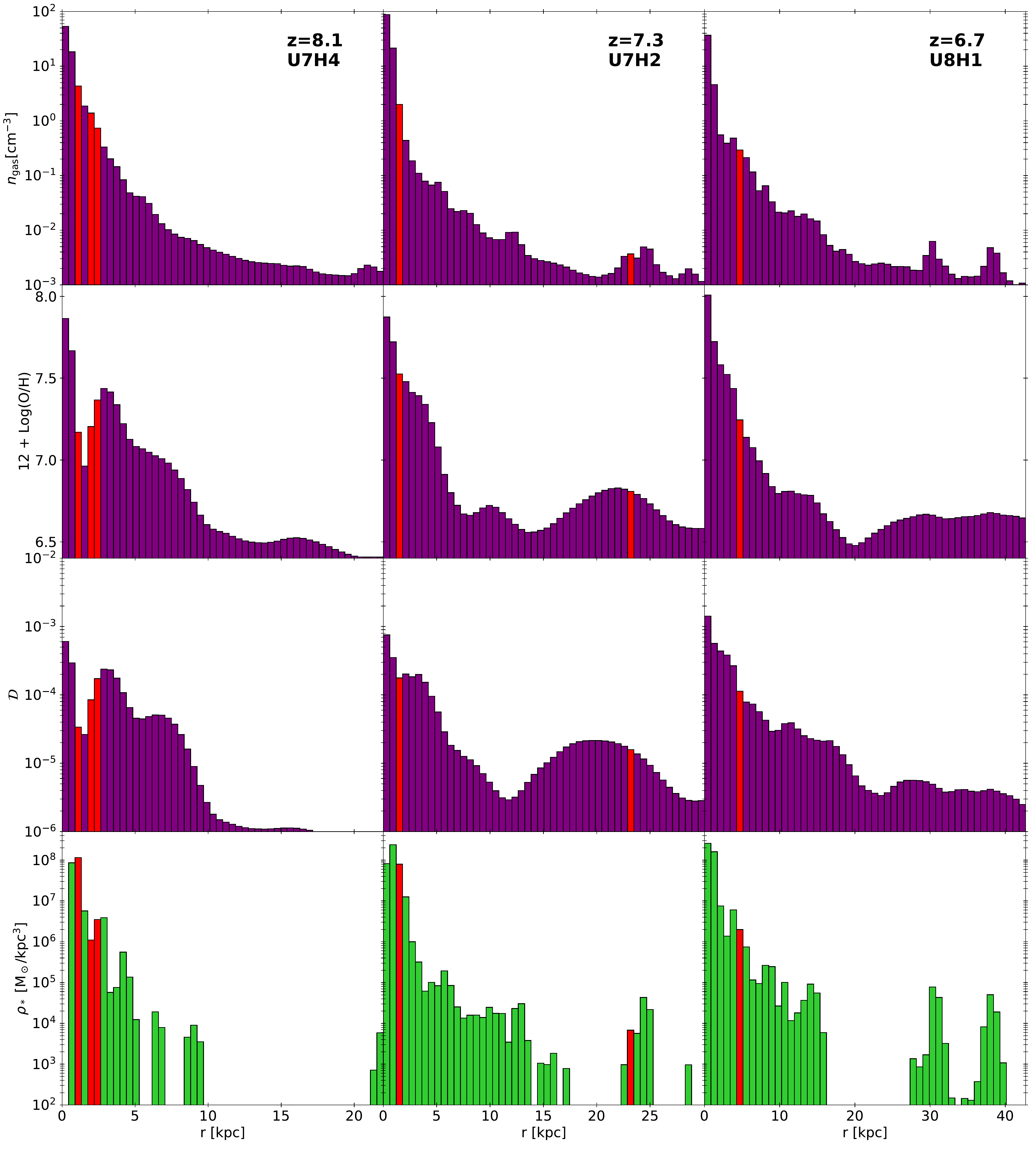}
    \caption{Radial profiles of the gas number density $n_\mathrm{gas}$ (\textbf{top row}), O/H ratio (\textbf{second row}), dust-to-gas ratio $\mathcal{D}$ (\textbf{third row}) and stellar mass density $\rho_\star$ (\textbf{bottom row}) are shown up to a distance of the order of the virial radius, for halos  U7H4 at $z = 8.1$, U7H2 at $z = 7.3$ and U8H1 at $z = 6.7$ (\textbf{left}, \textbf{middle} and \textbf{right columns} respectively). The shells containing Pop\ III stellar particles are highlighted in \textit{red}. Note that both the solar value $12 + \mathrm{Log (O/H)_\odot} = 8.69 \pm 0.05$ from \citet{Asplund_2009} and the local dust-to-gas ratio found by \citet{Zubko_2004} ($0.00568 \leq \mathcal{D} \leq 0.00813$, see their Table~6) lie well above the radial profiles shown in the second/third rows.}   \label{fig:radialProfiles_plotMatrix}
\end{figure*}

The position of Pop\ III stars on the maps in Figure~\ref{fig:snap020_U7H2_340.0_collapsedProjection_plotMatrix} show that these stellar systems are embedded in a highly inhomogeneously enriched environment. Therefore, the observability of their emitted spectra, including nebular emission lines, will strongly depend on the specific line-of-sight to the source (see Section \ref{sec:results_LOS}). The ionizing emission of Pop\ II binary systems, likely present in low-metallicity environments during the EoR (see e.g. \citealt{Stanway_Elridge_2018}), could also potentially mask the contribution of Pop\ III stars to the He$^+$-ionizing spectrum, making their identification less straightforward.

Table~\ref{tab:PopIII_localEnvironments} summarises the properties of active Pop\ III stellar populations found in the three selected halos U7H4 ($z = 8.1$), U7H2 ($z = 7.3$) and U8H1 ($z = 6.7$). Their metallicity spans a broad range of values ($-9.6 \leq \mathrm{Log}(Z_\mathrm{III} / \si{Z_{\odot}}) \leq -4.1$), and in one case we even find a pristine population (U7H2P2, with $Z_\mathrm{III} = 0$). For each of the three halos, the properties of the local environment surrounding Pop\ III stellar particles are determined by projecting the SPH kernel-weighted contribution of gas particles found within a distance of 85.9 pc, 118 pc and 167 pc, respectively. On these scales, the gas densities are much lower than those which characterize the birth clouds ($n_{\rm gas} \gtrsim 300 ~ \si{cm^{-3}}$). Yet, their values and those of the O/H and dust-to-gas ratio $\mathcal{D}$ reflect the typical environments where Pop\ III stars have formed, closer to the centre or in the outskirts of the galaxy (see also the maps in Figure~\ref{fig:sliceCuts_plotMatrix}).

\begin{table}
    \centering
    \caption{Properties of the Pop\ III stellar populations hosted by halos U7H4 ($z = 8.1$), U7H2 ($z = 7.3$) and U8H1 ($z = 6.7$) and of their local environment. We report the stellar particle ID, the stellar particle metallicity $\mathrm{Log} (Z_\mathrm{III} / \si{Z_\odot})$ and the gas number density  $n_\mathrm{gas} ~ [\si{cm^{-3}}]$, oxygen abundance $12 + \mathrm{Log} (\mathrm{O/H})$, and dust-to-gas mass ratio $\mathrm{Log} (\mathcal{D})$ in the unresolved region (i.e. 85.9~pc, 118~pc and 167~pc respectively) containing the stellar particle.}
    \begin{tabular}{ccccc}
        \hline
        part. ID & $\mathrm{Log} Z_\mathrm{III}$ & $\mathrm{Log} n_\mathrm{gas}$ & $12 + \mathrm{Log(O/H)}$ & $\mathrm{Log} \mathcal{D}$ \\       
        \hline
        U7H4P1 & -5.1 & 1.0 & 6.6 & -5.6 \\
        U7H4P2 & -4.1 & 0.5 & 6.2 & -6.2 \\
        U7H4P3 & -9.6 & 0.3 & 6.7 & -5.3 \\
        U7H4P4 & -5.1 & 1.0 & 7.0 & -4.9 \\
        \hline
        U7H2P1 & -4.5 & -0.2 & 7.7 & -3.3 \\
        U7H2P2 & $-\infty$ & -1.4 & 4.0 & -8.5 \\
        \hline
        U8H1P1 & -4.8 & 0.1 & 7.3 & -3.6 \\
        \hline
    \end{tabular}
    \label{tab:PopIII_localEnvironments}
\end{table}

Figure \ref{fig:radialProfiles_plotMatrix} provides a complementary view of the Pop\ III halo environments by showing spherically-averaged radial profiles of $n_\mathrm{gas}$, metallicity (in units $12 + \mathrm{Log}(\mathrm{O/H})$), dust-to-gas ratio ($\mathcal{D}$) and stellar mass density ($\rho_\star$), as a function of the radial distance $r$ from the centre of mass of the system in physical kpc units. While the angular distribution of the above quantities is smoothed by the spherical average, these profiles are useful to appreciate global gradients as a function of $r$ and provide a first clue of the properties along an average line-of-sight to the centre of the systems (see next section for more details). In all panels, the radial distance of Pop\ III stars is indicated with vertical red bars.

As expected, all the profiles show a global negative gradient from the centre to the virial radius. However, the decline is not continuous and structures emerge even at $r \sim 5 - 10$~kpc and at $r \gtrsim 20 -30$~kpc, especially for U8H1 at $z = 6.7$ and U7H2 at $z = 7.3$, which are in a more advanced phase of their assembly, compared to U7H4 at $z = 8.1$ (see also Section \ref{sec:results_PopIIIhalosSFH}). As an example, although the gas number density drops by three orders of magnitude within $r\sim 5$~kpc from the center, star-forming clumps are present also at larger distances (compare top and bottom rows), confirming the qualitative picture described by Figures~\ref{fig:sliceCuts_plotMatrix} and \ref{fig:snap020_U7H2_340.0_collapsedProjection_plotMatrix} of a complex assembly scenario occurring in DM halos hosting galaxies with Log(M$_\star/\rm M_{\odot}) \geq 9.0$ at these redshifts.

The radial profiles of the oxygen abundance and dust-to-gas mass ratio appear to trace each other, reflecting the stellar mass density distribution in the halos. However, there are some notable deviations, with the dust radial profiles showing in general a broader distribution around stellar overdensities and a smoother negative gradient (see however the case of U7H4 at $z = 8.1$, where the dust-to-gas mass ratio appears to be more centrally concentrated than the oxygen abundance). The irregular internal structure characterising these systems have an important impact on the emerging spectra, which will be strongly dependent on the line-of-sight to the source.

\begin{figure*}
    \centering
    \includegraphics[width=\linewidth]{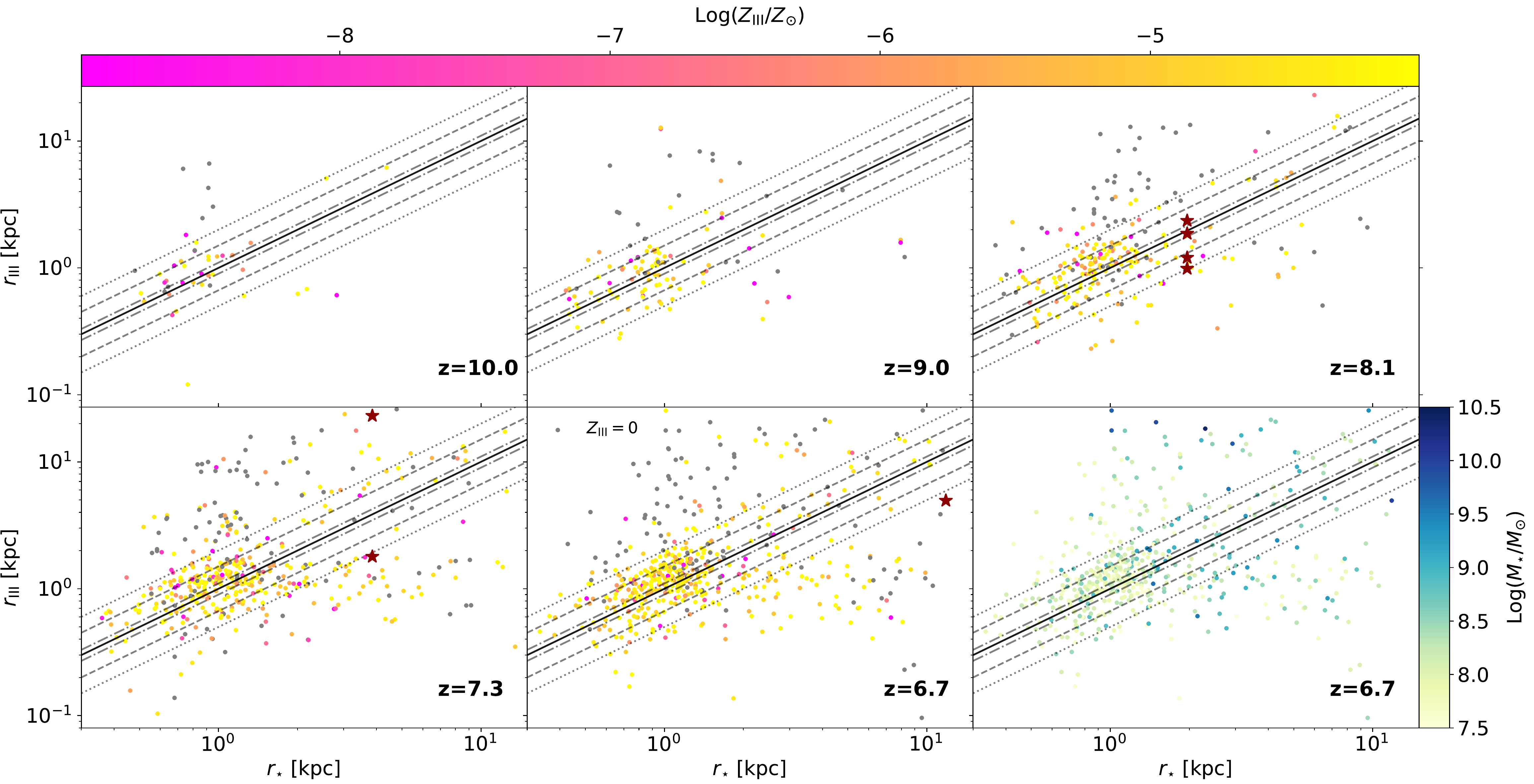}
    \caption{Distance $r_\mathrm{III}$ of Pop III stellar particles from the center of mass of their hosting DM halos as a function of the stellar mass-weighted radius $r_\star$. Results are shown for all Pop IIIs found in resolved halos (Log$(M_\star / \si{M_\odot}) \geq 7.5$) of the simulated cubes U6, U7, U8, U10, U12, U13 at redshifts $z = 10.0, 9.0, 8.1, 7.3, 6.7$ (\textbf{first} to \textbf{second-last panels}). The metallicity $Z_\mathrm{III}$ of the Pop III stellar particles is shown in \textit{pink-yellow scale} (with zero-metallicity particles - out of this scale - in \textit{grey}), while the \textbf{last panel} replicates the panel at $z = 6.7$ with Pop III particles in \textit{yellow-blue scale} according to the stellar mass $M_\star$ of their host halos. Black, straight lines indicate where $r_\mathrm{III} = tol \cdot r_\star$, for different values of the tolerance $tol$: $tol = 1$ (\textit{solid lines}), $tol = 0.9 - 1.1$ (\textit{dashed-dotted lines}), $tol = 1/1.5 - 1.5$ (\textit{dashed lines}) and $tol = 0.5 - 2$ (\textit{dotted lines}). The Pop III stellar particles found in halos U7H4, U7H2 and U8H1 (see Table~\ref{tab:PopIII_localEnvironments}) are highlighted by \textit{dark-red stars}.}
    \label{fig:popIIIDistancesFromCM}
\end{figure*}

To further explore the statistics of Pop\ III-forming environments across the whole sample, Figure~\ref{fig:popIIIDistancesFromCM} shows the distance from the centre of mass of their host halos, $r_\mathrm{III}$, of all Pop III stellar particles in resolved halos at $z = 10.0, ~ 9.0, ~ 8.1, ~ 7.3$ and 6.7 as a function of the stellar mass-weighted radius $r_\star$. We hereafter define ``external Pop III'' those Pop\ IIIs found at $r_\mathrm{III} > tol \cdot r_\star$, and explore the results for different values of the tolerance $tol$ between 1 and 2. When $tol = 1$ more than 70\% of Pop\ IIIs are external, while this fraction drops to 55\% and 47\% when $tol = 1.5$ and $2$, respectively. As a reference, the stellar particles of Table~\ref{tab:PopIII_localEnvironments} are marked by dark-red stars in the plots. While the Pop\ IIIs of halos U7H2 and U8H1 are clearly classified as internal/external according to the $r_\star$-criterion, those of U7H4 are all found within a distance of the order of two times $r_\star$.

We also find that the fraction of external Pop\ IIIs is rather insensitive to redshift $z$, to the age of Pop III stellar particles $t_\mathrm{III}$ and to the stellar mass of their host halo $M_\star$, although the rare massive Pop III-forming galaxies (Log$(M_\star / \si{M_\odot}) \geq 9.5$) all seem to be forming Pop\ IIIs at their periphery rather than in the central regions (see e.g. the last panel of Figure~\ref{fig:popIIIDistancesFromCM}). By looking at the metallicity distribution of Pop III particles ($Z_\mathrm{III}$, first to second-last panels of Figure~\ref{fig:popIIIDistancesFromCM}), we find instead that the majority of pristine populations (Log$(Z_\mathrm{III} / \si{Z_\odot}) < -9$, in grey) are found in the external regions, i.e. in pristine clouds at the periphery. This is expected, as these regions are more likely to be preserved from metal pollution arising from the high-density, star-forming regions near the centre and/or accrete pristine gas from the large scale.

\subsection{Environments across different lines-of-sight}
\label{sec:results_LOS}
In this section we study the variety of environments encountered by different Lines-Of-Sight (LOS) directed towards Pop\ III stars. In fact, their observability and identification will heavily depend on the properties of emitters and absorbers along the LOS.
Hence, the expected scatter of gas and dust properties across different LOS provides indications on the range of optical depths 
encountered by stellar and nebular emission.

We are particularly interested in the dust component, as dust grains are the most important absorbers for the HeII$\lambda$1640 line, whose properties provide key spectral diagnostics to identify Pop\ III stars \citep{Inoue_2011, Nakajima_Maiolino_2022, Katz_2022, Saxena_2020b, Saxena_2020a}. Models of dust mixtures reproducing the extinction observed in the Milky Way\footnote{\url{https://www.astro.princeton.edu/~draine/dust/dust.html}} \citep{Weingartner_Draine_2001, Li_Draine_2001, Draine_2003_interstellarDust, Draine_2003_dustScatteringUV, Draine_2003_dustScatteringX, Glatzle_2019} predict indeed a non-negligible dust absorption cross section per unit dust mass\footnote{By comparison, the peak value is $\sigma_\mathrm{a} / m_\mathrm{dust} \simeq 8 - 15 \times 10^4 ~ \si{cm^2.g^{-1}}$ (at $\lambda \simeq 723 ~ \si{\angstrom}$), see for instance Figure 1 of \citet{Glatzle_2019}.} at $\lambda = 1640 ~ \si{\angstrom}$ ($\sigma_\mathrm{a} / m_\mathrm{dust} \gtrsim 2.5 \times 10^4 ~ \si{cm^2.g^{-1}}$). Recently, \citet{Curtis-Lake_2023} failed to detect the HeII line in four metal-poor galaxies at $z > 10$, yielding $2\sigma$ upper limits on the HeII equivalent widths (EW) of $\simeq 6 - 15.4 ~ \si{\angstrom}$. Although the authors argued these measured limits are not yet particularly constraining, the lack of detectable emission lines might be ascribed to high levels of dust absorption; indeed, two of the objects do indicate moderate levels of dust (V-band optical depth, $\tau_\mathrm{V} \sim 0.2$), albeit with large uncertainties. \citet{Roberts-Borsani_2022} also reported no prominent emission lines arising from the spectrum of a galaxy observed with JWST/NIRSpec at $z = 9.76$.

We continue to focus on the prototypical halo U7H2 at $z = 7.3$. As explained in Section \ref{sec:results_PopIIIhalosSFE}, one of the 
Pop\ III stellar populations of the halo (U7H2P1) is found in a central region, at the periphery of a Pop\ II-dominated clump; the other (U7H2P2) is in an isolated, pristine gas cloud, far from the centre. To statistically study the gas/dust distribution, the halo is sampled by shooting a million random LOS originating respectively from U7H2P1 and U7H2P2. These would correspond to the path of photons travelling on a straight line (i.e. without scatterings) in random directions starting from the Pop III sources, until they finally escape the halo (see \citealt{Graziani_2018} for more details). We collectively refer to these two groups of LOS as ``LOSP1'' and ``LOSP2''.

\begin{table}
    \centering
    \caption{Total distance $d_\mathrm{tot}$~[kpc], gas column density $\mathrm{Log} (N_\mathrm{gas}/\si{cm^{-2}})$, oxygen surface density $\mathrm{Log} [\Sigma_\mathrm{O}/(\si{M_\odot kpc^{-2}})]$, dust surface density $\mathrm{Log} [\Sigma_\mathrm{dust}/(\si{M_\odot kpc^{-2}})]$ and optical depth at $\lambda = 1640 ~ \si{\angstrom}$ Log$\tau$ crossed by the LOS originating from the Pop\ III near the centre (LOSP1) and from the peripheral Pop\ III (LOSP2) respectively in Halo U7H2 at $z = 7.3$. $\tau$ is computed assuming a dust cross section per unit dust mass $\sigma_\mathrm{a} / m_\mathrm{dust} \simeq 3.8 \times 10^4 ~ \si{cm^2.g^{-1}}$ at $\lambda = 1640 ~ \si{\angstrom}$ (for an inter-stellar reddening parameter $R_\mathrm{V} = 3.1$). Only the LOS crossing the maximum, mean, median and minimum $\Sigma_\mathrm{dust}$ are displayed.}
    \begin{tabular}{lccccc}
        \hline
         & $d_\mathrm{tot}$ & $\mathrm{Log} \, N_\mathrm{gas}$ & $\mathrm{Log} \, \Sigma_\mathrm{O}$ & $\mathrm{Log} \, \Sigma_\mathrm{dust}$ & $\mathrm{Log} \tau$ \\
          \hline
         \multicolumn{6}{c}{LOSP1} \\
         \hline
         Max. & 39.9 & 23.7 & 6.43 & 6.37 & 1.27 \\
         Mean & 31.9 & 22.6 & 5.16 & 4.90 & -0.20 \\
         Median & 29.1 & 21.4 & 3.84 & 3.87 & -1.23 \\
         Min. & 40.5 & 21.0 & 3.34 & 3.23 & -1.87 \\
         \hline
         \multicolumn{6}{c}{LOSP2} \\
         \hline
         Max. & 54.4 & 23.8 & 6.46 & 6.40 & 1.30 \\   
         Mean & 50.8 & 21.0 & 2.50 & 2.59 & -2.51 \\
         Median & 40.5 & 20.1 & 0.51 & -1.12 & -6.22 \\
         Min. & 30.4 & 20.7 & -1.67 & -3.10 & -8.20 \\
         \hline
    \end{tabular}
    \label{tab:LOS}
\end{table}
\begin{figure*}
    \centering
    \includegraphics[width=0.95\textwidth]{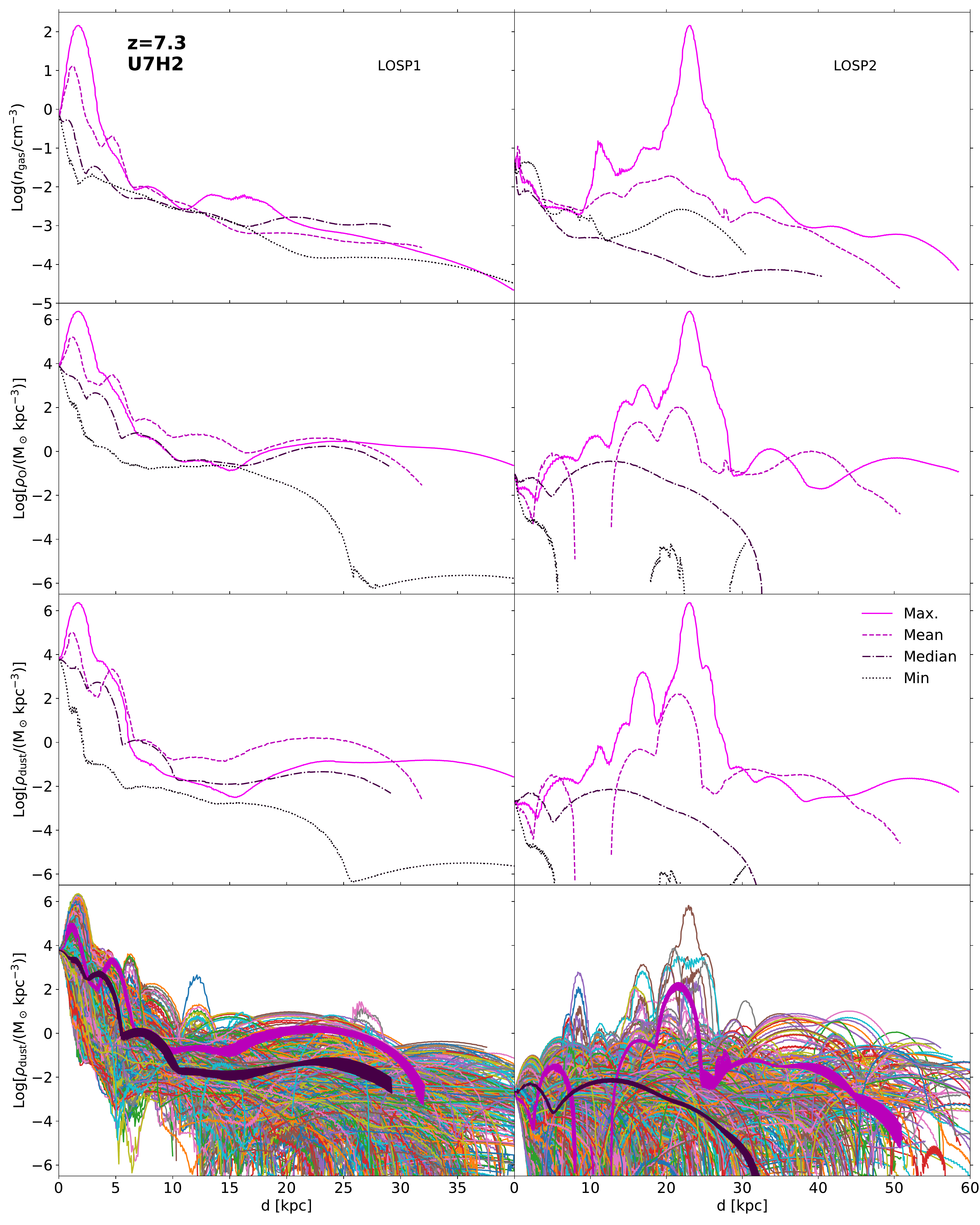}
    \caption{Profiles of the gas number density $n_\mathrm{gas}$ (\textbf{top row}), oxygen mass density $\rho_\mathrm{O}$ (\textbf{second row}) and dust mass density $\rho_\mathrm{dust}$ (\textbf{third row}) along LOS originating from the Pop\ III near the centre (LOSP1, \textbf{left panels}) and from the external Pop\ III (LOSP2, \textbf{right panels}) in Halo U7H2 at $z = 7.3$, as a function of the distance $d$ from the Pop\ III. The LOS crossing the maximum, mean, median and minimum dust surface density $\Sigma_\mathrm{dust}$ are respectively shown as \textit{solid}, \textit{dashed}, \textit{dashed-dotted} and \textit{dotted lines}, with progressively darker shades of \textit{purple}. \textbf{Bottom panels:} $\rho_\mathrm{dust}$ profile for a selection of 1000 random LOS. The superimposed \textit{light/dark-purple} regions show the dust density profile along all the LOS (243/302 for the LOSP1, 85/161 for the LOSP2) coming out of a square area made of $15 \times 15$ cells ($\sim 1.77 ~ \si{kpc} \times 1.77 ~ \si{kpc}$) around the mean/median LOS.}
    \label{fig:LOS}
\end{figure*}

The first six panels of Figure \ref{fig:LOS} show the profile of the gas number density ($n_\mathrm{gas}$), oxygen mass density ($\rho_\mathrm{O}$) and dust mass density ($\rho_\mathrm{dust}$) as a function of the distance $d$ along the LOS from each source (U7H2P1 in the left panels, U7H2P2 in the right panels). Results are shown for the LOS crossing the maximum, mean, median, and minimum dust surface density $\Sigma_\mathrm{dust}$. These values for each of the eight LOS are also listed in Table \ref{tab:LOS}, together with the total distance crossed ($d_\mathrm{tot}$), gas column density ($N_\mathrm{gas}$) and oxygen surface density ($\Sigma_\mathrm{O}$). The optical depth ($\tau$) at $\lambda = 1640 ~ \si{\angstrom}$ is also reported in the table; it is computed as $\tau = \sigma_\mathrm{a} / m_\mathrm{dust} \times \Sigma_\mathrm{dust}$, assuming $\sigma_\mathrm{a} / m_\mathrm{dust} \simeq 3.8 \times 10^4 ~ \si{cm^2.g^{-1}}$ at $\lambda = 1640 ~ \si{\angstrom}$ (for an inter-stellar reddening parameter $R_\mathrm{V} = 3.1$, \citealt{Glatzle_2019}). We warn the reader that $\tau$ is just an indication of the level of absorption induced by dust grains, while a more accurate determination of the signal removed from the LOS certainly requires to take into account the probability of scattering and the non-negligible deflection angles involved at this frequency\footnote{For example, assuming $R_\mathrm{V} = 3.1$, the values of the albedo and the scattering asymmetry parameter are $\omega \simeq 0.4$ and $\langle \cos\theta \rangle \simeq 0.6$ at $\lambda = 1640 ~ \si{\angstrom}$.}.

While the maximum $\Sigma_\mathrm{dust}$ in the LOSP1/LOSP2 cases is similar ($\mathrm{Log} [\Sigma_\mathrm{dust}/(\si{M_\odot kpc^{-2}})] \simeq 6.4$), the values of the mean/median/minimum LOSP1 are about two/five/six orders of magnitude higher with respect to the LOSP2. This is easily explained by looking at the dust profile along the LOS. The maximum LOS are both passing through the galaxy centre, where the highest values of the gas, metals and dust density are found (see the peaks at $d \sim 1.5 ~ \si{kpc}$ for the LOSP1 and $d \sim 23 ~ \si{kpc}$ for the LOSP2). The mean, median and minimum LOS are instead travelling through very different environments in the LOSP1/LOSP2 cases. Also note that $\Sigma_\mathrm{O}$ is about the same order of $\Sigma_\mathrm{dust}$, showing the metals and dust distributions have a strong spatial correlation. By comparison, the variation of $N_\mathrm{gas}$ between the LOSP1 and the LOSP2 cases is much lower. Indeed, the gas is more uniformly distributed with respect to metals and dust.

These results clearly demonstrate that the observability of Pop\ III stars, embedded in high-$z$ dusty galaxies, further depends on the inclination of the galaxy with respect to the observer. In the ``best-case'' scenario, the photons emitted by a Pop\ III star-forming region encounter almost no dust along their path ($\mathrm{Log} [\Sigma_\mathrm{dust}/(\si{M_\odot kpc^{-2}})] \lesssim -3$) and are expected to experience a low degree of absorption. In the ``worst-case'' scenario, on the other hand, they come across a significant dust screen ($\mathrm{Log} [\Sigma_\mathrm{dust}/(\si{M_\odot kpc^{-2}})] \gtrsim 6$) and hence will be heavily absorbed. The corresponding optical depth would be $\tau \lesssim 10^{-8}$ and $\tau \gtrsim 10$, respectively for the two cases.

\begin{figure*}
    \centering
    \includegraphics[width=\linewidth]{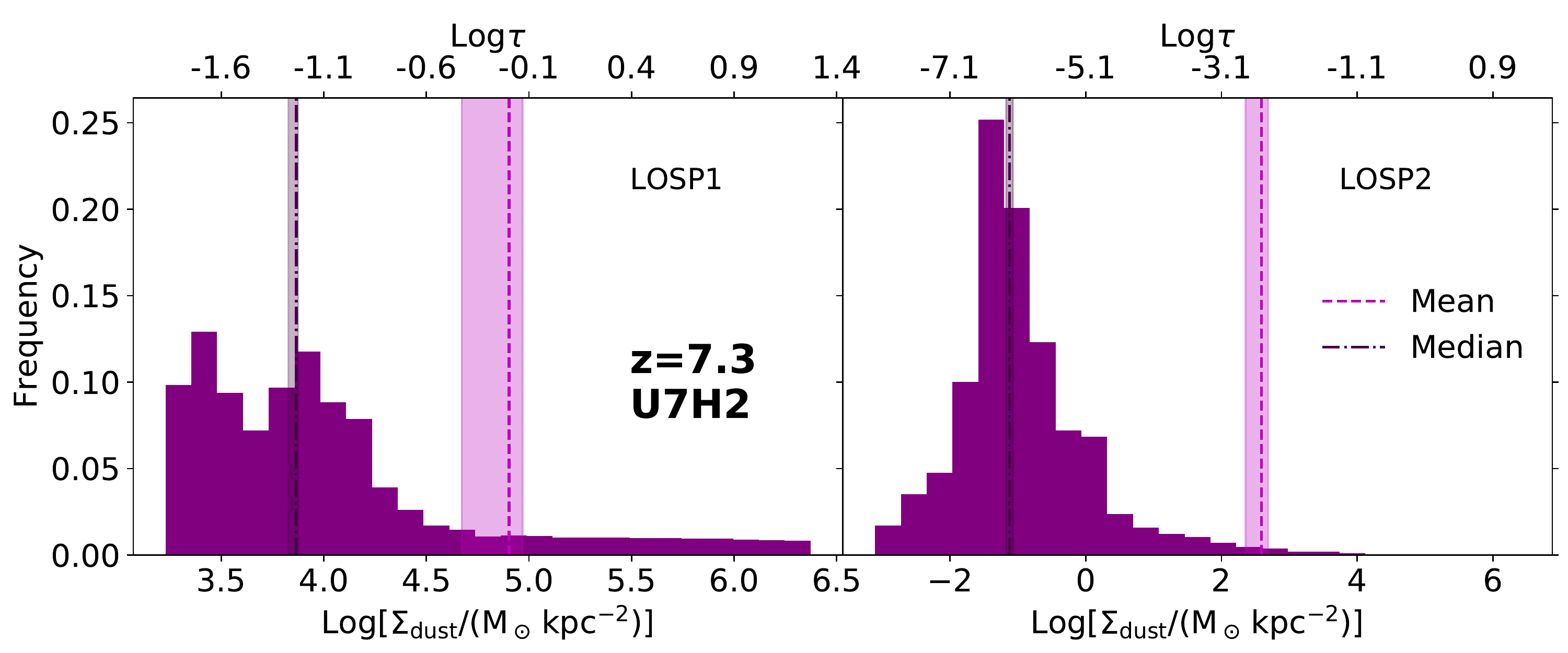}
    \caption{Histogram of the relative frequencies of the dust surface densities $\Sigma_\mathrm{dust}$ crossed by LOS originating from the Pop\ III near the centre (LOSP1, \textbf{left panels}) and from the external Pop\ III (LOSP2, \textbf{right panels}) in Halo U7H2 at $z = 7.3$. The corresponding values of $\tau$, computed assuming a dust cross section per unit dust mass $\sigma_\mathrm{a} / m_\mathrm{dust} \simeq 3.8 \times 10^4 ~ \si{cm^2.g^{-1}}$ at $\lambda = 1640 ~ \si{\angstrom}$ (for an inter-stellar reddening parameter $R_\mathrm{V} = 3.1$), are displayed in the top axis. The superimposed \textit{light/dark-purple} shaded regions show the min-max spread of the dust surface density profile along all the LOS coming out of a square area made of $15 \times 15$ cells around the mean/median LOS (refer to the bottom panel of Figure \ref{fig:LOS}).}
    \label{fig:LOSDustDensityDistribution}
\end{figure*}

The bottom panels of Figure \ref{fig:LOS} show the $\rho_\mathrm{dust}$ profile for a selection of a thousand random LOS to further demonstrate the huge scatter existing in the entire sample. We see the scatter can reach values up to almost 12~dex at fixed distance. This strong variability is a direct consequence of the irregular and clumpy nature of the halo, that has been extensively commented in Section \ref{sec:results_PopIIIhalos}.

To provide useful hints for observations, in the same panels we have also compared the global scatter traced by our projected grid to the one detected among LOS coming out of an area with size comparable to JWST/NIRSpec shutters. The shutters have an angular size in the sky of $0.20 ~ \si{arcsec} \times 0.46 ~ \si{arcsec}$ \citep{Ferruit_2022}, corresponding to a physical size of $1.04 ~ \si{kpc} \times 2.39 ~ \si{kpc}$ at $z = 7.3$ with our assumed cosmology. By comparison, our grid cells have a physical side of $\simeq 118 ~ \si{pc}$, meaning a NIRSpec shutter would be covered by approximately 180 of our cells. Here we considered a slightly bigger area of $15 \times 15$ cells, i.e. $\simeq 1.77 ~ \si{kpc} \times 1.77 ~ \si{kpc}$, around the exit point of the LOS along which $\Sigma_\mathrm{dust}$ has a value close the mean/median (light/dark-purple regions). A total of 243/302 and 85/61 LOS respectively are found in this area for the LOSP1 and the LOSP2 cases. The scatter among these LOS is always lower than 1~dex at fixed distance. This means, in turn, that the environments crossed by photons that would fall into a single NIRSpec/MOS resolution element are not expected to vary more than this threshold, at least for a source similar to U7H2.

To further appreciate the scatter found among these two groups of LOS, we also show their $\Sigma_\mathrm{dust}$ min-max spread in Figure \ref{fig:LOSDustDensityDistribution} (light/dark-purple shaded regions), with respect to the $\Sigma_\mathrm{dust}$ distribution across the whole sample. By looking at the histogram of the relative frequencies of $\Sigma_\mathrm{dust}$ for the LOSP1 (left panel) and the LOSP2 (right panel) it is evident that in both cases the distribution is peaked at low dust surface densities ($\mathrm{Log} [\Sigma_\mathrm{dust}/(\si{M_\odot kpc^{-2}})] \sim 3-4$ for the LOSP1, $ \sim -1$ for the LOSP2), although a long tail of high-surface-density LOS (up to $\mathrm{Log} [\Sigma_\mathrm{dust}/(\si{M_\odot kpc^{-2}})] \sim 6$) is also present. This means that LOS crossing relatively low amounts of dust are more likely, especially in the case of the ones towards the external Pop III, given its positioning.

The LOSP2 cases also span a much larger range in $\Sigma_\mathrm{dust}$ ($\sim 10$~dex) with respect to the LOSP1 ($\sim 3$~dex). By comparison, the LOS that would fall in an area of the order of NIRSpec shutters only span $\sim 0.3$~dex around the mean. An even lower range of $\simeq 0.1$~dex is also found around the median LOS, for both the LOSP1 and the LOSP2. The above analysis confirms that line-of-sights selected by the shutters have a variability significantly reduced with respect to the whole sample.

\section{Conclusions}
\label{sec:conclusions}

Motivated by a significant collection of recent numerical models suggesting a late Pop\ III star formation during the Epoch of Reionization, in this paper we investigated the formation of Pop\ III stars during EoR at different scales: on cosmological boxes of $50h^{-1}$~cMpc/side and in galaxies with Log(M$_\star/\rm M_{\odot}) \gtrsim 9.0$. With this aim in mind we also addressed the following key questions: (i) how does the Pop\ III - Pop\ II SFR evolve through cosmic times? (ii) Which classes of galaxies are mainly hosting Pop\ III stars during the EoR? (iii) In which environments are Pop\ III stars embedded within the most massive galaxies?\\
Our exploration is performed by adopting the large statistics provided by recent \texttt{dustyGadget} simulations, from which we extracted an interesting sample of galaxy candidates already accessible to JWST observations. The properties of such targets and their star-forming regions in terms of stellar/gas/dust components and SFHs were also investigated in details.

Hereafter we summarize our main conclusions: 
\begin{enumerate}
    \item A non-negligible, late Pop\ III star formation ($\Psi_\mathrm{III}  \sim 10^{-3.4} - 10^{-3.2} ~ \si{M_\odot.yr^{-1}.cMpc^{-3}}$) is still occurring down to the end of the EoR ($z \sim 6 - 8$), at redshifts that are well within the reach of deep photometric and spectroscopic surveys with JWST. This is in agreement with previous models, despite the very different adopted simulation strategies and included physics, and despite the large scatter between all the various predictions.
    \item Resolved Pop\ III galaxies ($M_\star \gtrsim 3 \times 10^7 ~ \si{M_\odot}$) are mainly found on the MS, at high SFRs. Their high SFRs may be induced by a sustained rate of accretion of pristine gas from the large scale, triggering Pop\ III star formation at later times. For some of the halos, this scenario is also supported by their increasing SFHs and by their preferred positioning in high-density regions of the cosmic web ($n_\mathrm{gas}/\overline{n}_\mathrm{gas} \gtrsim 1000$). Note a more thorough statistical analysis would be required to further confirm this hypothesis. 
    \item Even in some of the most massive galaxies which are already hosting a persistent Pop\ II star formation, Pop\ III stars can survive, potentially allowing their direct detection. At $z \sim 6 - 7$, around $10 - 50 \%$ of the rare massive galaxies with $M_\star \gtrsim 3 \times 10^9 ~ \si{M_\odot}$ may host Pop\ III stars, although with a low Pop\ III/Pop\ II relative stellar mass fraction $\lesssim 0.1 \%$.
    \item In this class of objects, Pop\ III stars are found both in the outskirts of metal-enriched regions ($12 + \mathrm{Log (O/H)} \sim 6.2 - 7.7$, $\mathrm{Log} \mathcal{D} \sim - (6.2 - 3.6)$) and in isolated, pristine gas clouds (e.g. $12 + \mathrm{Log (O/H)} \sim 4.0$, $\mathrm{Log} \mathcal{D} \sim -8.5$). In this latter case, their signal may be less contaminated by other stellar populations.
    \item By exploring the environments crossed by a million random LOS, we found that the amount of dust through the various LOS strongly depends on the inclination of the target galaxy and its Pop\ III-forming environments with respect to the plane of the observer. Indeed, photons can travel through columns of dust that go from $ \Sigma_\mathrm{dust} \lesssim 10^{-3} ~ \si{M_\odot kpc^{-2}}$ up to $ \Sigma_\mathrm{dust} \gtrsim 10^{6} ~ \si{M_\odot kpc^{-2}}$.
\end{enumerate}

Further investigations are certainly required to make any claim on the detectability of Pop\ III stars during EoR as we still need to study (i) the level of confusion of Pop\ III signals from nearby Pop\ II stellar populations, and (ii) how much of their intrinsic flux is absorbed by the ISM of the hosting galaxies. Analyses of the stellar continuum and nebular emission, and RT simulations performed on the selected targets will provide a more thorough insight on these key questions. 

\section*{Acknowledgments}
We would like to thank the Referee, Liu Boyuan, for his insightful comments and suggestions. LG, RS and KO acknowledge support from the Amaldi Research Center funded by the MIUR program "Dipartimento di Eccellenza" (CUP:B81I18001170001). KO also acknowledges support from the Japan Society for the Promotion of Science by Grants-in-Aid for Scientific Research (17H06360, 17H01102, 22H00149). We have benefited from the publicly available programming language \texttt{Python}, including the \texttt{numpy}, \texttt{matplotlib} and \texttt{scipy} packages.

\section*{Data Availability}
The data underlying this article will be shared on reasonable request to the corresponding author.

\bibliographystyle{mn2e}
\bibliography{main}

\begin{appendix}
\section{SFRD comparison with small-scale simulations}
\label{sec:AppSFRD}

\begin{figure}
    \centering
    \includegraphics[width=\linewidth]{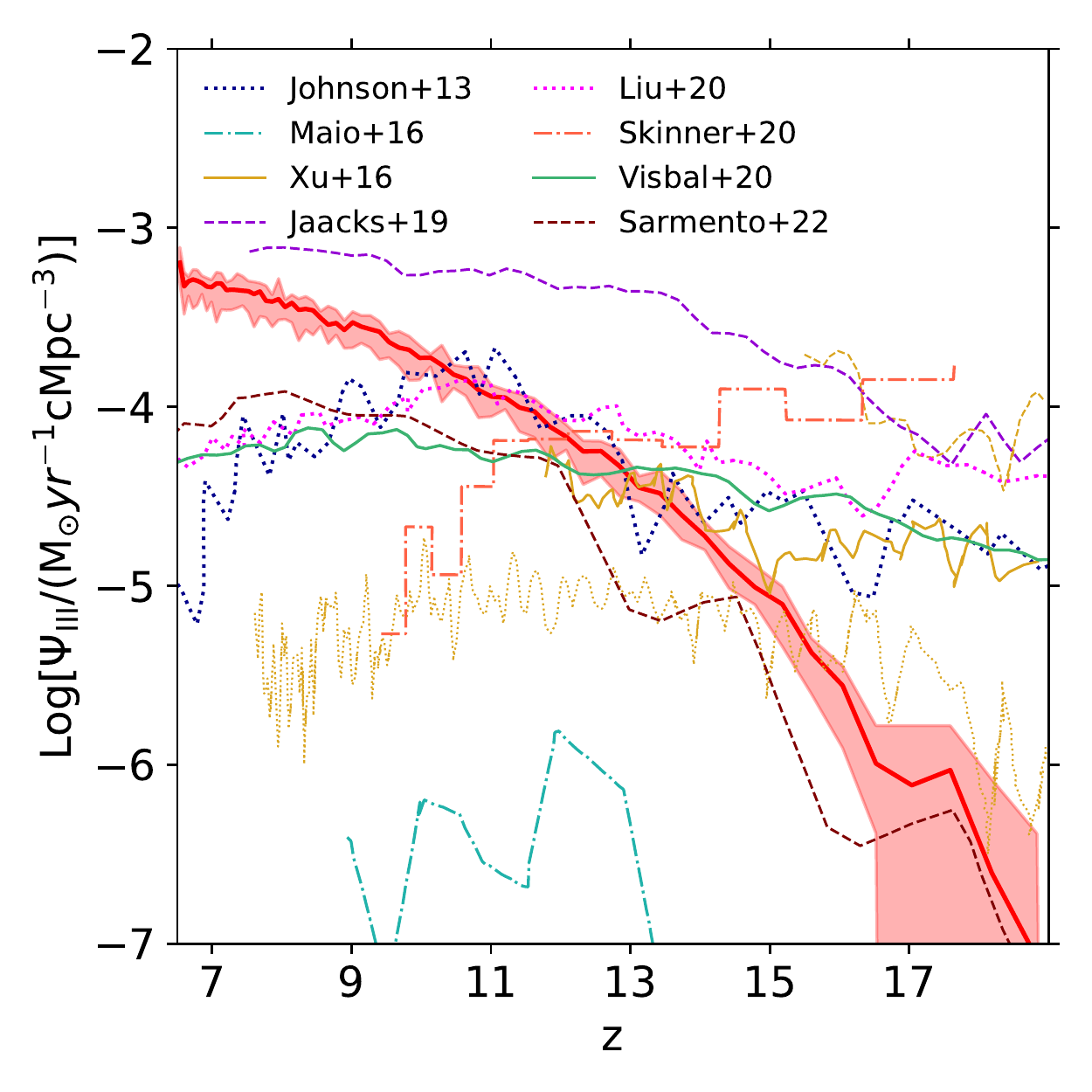}
    \caption{Average Pop\ III SFRD $\Psi_\mathrm{III}$ (computed as in the left panel of Figure \ref{fig:SFRD_pop_av+obs+U12}) compared with the results of other small-scale (box sizes $\lesssim 4h^{-1} ~ \si{cMpc}$) models and simulations, i.e. \citet{Johnson_2013} (\textit{dark-blue, dotted line}),  \citet{Maio_2016} (\textit{light-blue, dashed-dotted line}), \citet{Xu_2016_XRB} (\textit{gold lines} - the ``Normal'', ``Void'' and ``Rarepeak'' simulations are shown respectively in \textit{solid}, \textit{thin-dotted} and \textit{thin-dashed} linestyle), \citet{Jaacks_2019} (\textit{purple, dashed line}), \citet{Liu_Bromm_2020_2} (\textit{hotpink, dotted line}), \citet{Skinner_Wise_2020} (\textit{orange, dashed-dotted line}), \citealt{Visbal_2020} (\textit{green, solid line}) and \citet{Sarmento_Scannapieco_2022} (\textit{brown, dashed lines}). See Table \ref{tab:PopIII_models} for a compilation of the models used as a comparison here and in Figure \ref{fig:SFRD_pop_av+modLargeScale}, with their main features.}
    \label{fig:SFRD_pop_av+modSmallScale}
\end{figure}

Here we compare our predictions for the SFRD to a wider set of simulations collected in Table \ref{tab:PopIII_models}. Data is sorted by adopted box side length in order to separate simulations performed on more than ten cMpc scales from smaller boxes. For each study, we provide in Table \ref{tab:PopIII_models} the reference paper, code name, adopted numerical scheme\footnote{For hydrodynamical simulations, these are either particle-based (SPH, e.g. \citealt{Springel_2005_Gadget-2, Springel_2008_Gadget-3}), grid-based Adaptive Mesh Refinement (AMR, e.g. \citealt{Rosdahl_2013}) or Meshless Finite Mass (MFM, e.g. \citealt{Hopkins_2018_FIRE-2}) schemes. Two Semi-Analytical Models (SAM, i.e. Trinca et al. in prep., \citealt{Visbal_2020}) are also included in the list.}, box side length and mass of DM particles. Figure~\ref{fig:SFRD_pop_av+modSmallScale} compares the predicted $\Psi_{\rm III}(z)$ resulting from small-scale/high-mass-resolution simulations ($L \lesssim 4h^{-1}$~cMpc, see second group in Table \ref{tab:PopIII_models}); a comparison with large-scale simulations ($L \geq 10h^{-1}$~cMpc, first group in Table \ref{tab:PopIII_models}) has been already presented in Figure~\ref{fig:SFRD_pop_av+modLargeScale}.

The summary in Table \ref{tab:PopIII_models} shows that recent studies have adopted different numerical strategies and physical assumptions, and explore Pop\ III formation at different scales. Predictions of $\Psi_{\rm III}(z)$ at scales $L \geq 10h^{-1}$~cMpc (i.e. more similar to the one of \texttt{dustyGadget}) are shown in Fig. \ref{fig:SFRD_pop_av+modLargeScale}. These correspond to the results by \citet{Pallottini_2014}, \citet{Sarmento_2018} and by the semi-analytical model \texttt{CAT} (Trinca et al. in prep.); even though the last model does not have a proper definition of simulated volume, it spans a large number of Pop\ III star-forming environments, representative of a large-scale cosmological volume. Also note that the quoted $\geq 10h^{-1}$~cMpc simulations differ by a factor $ > 100$ in the simulated volumes, and by a remarkably different DM particle mass. 

In terms of radiative feedback, all large-scale models adopt a global ionizing UV background. The semi-analytic code \texttt{CAT} also accounts for feedback from a LW background. Apart from studies based on \texttt{GIZMO} \citep{Jaacks_2019, Liu_Bromm_2020_2} and the FiBY simulations \citep{Johnson_2013}, which implement a small-scale LW and UV background either with or without a semi-analytic treatment of local sources and shielding, all the other small-scale hydrodynamical simulations adopt a combination of a background with proper RT from stellar populations, at least for photons in the LW/UV bands\footnote{\citet{Maio_2016}, for example, follows local radiation from 150 frequency bins for photons originated by stars according to Pop\ III and Pop\ II Spectral Energy Distributions (SED).}. All hydrodynamical simulations but \citet{Pallottini_2014} implement primordial chemistry. With the exception of simulations based on the \texttt{ENZO} code (explicitly including an evaluation of the H$_2$ fraction in their star formation recipe), all models adopt an H-based star formation recipe.

Large differences are also found in the implementation of chemical feedback: \cite{Johnson_2013}, \cite{Maio_2016}, \texttt{dustyGadget} simulations and \texttt{CAT} semi-analytic model create metals from stars by adopting mass and metallicity-dependent yields and follow single atomic species during metal pollution and spreading; \texttt{RAMSES}, \texttt{ENZO} and \texttt{GIZMO}-based methods derive instead the metal mass through an average yield. Moreover, apart from \texttt{dustyGadget} and \texttt{CAT}, no models account for cosmic dust.

\begin{table}
    \centering
    \caption{Independent theoretical models accounting for Pop\ III star formation. Table columns show: model reference, code name, type of adopted numerical scheme, box size $L$~$[\si{cMpc}/h]$, adopted DM particle mass $m_\mathrm{DM}$~$[10^5 ~ \si{M_\odot}/h]$. All the listed theoretical models are compared in Figures \ref{fig:SFRD_pop_av+modLargeScale} and \ref{fig:SFRD_pop_av+modSmallScale}. See Figure \ref{fig:SFRD_pop_av+modSmallScale} for references.}
    \begin{tabular}{l|c|c|c|c}
         & Code & Scheme & $L$ & $m_\mathrm{DM}$ \\
        \hline
      This work & \texttt{dustyGadget} & SPH & 50 & 353.00 \\
      Sarmento+18 & \texttt{RAMSES} & AMR & 12 & 0.99 \\
      Pallottini+14 & \texttt{RAMSES} & AMR & 10 & 4.70 \\
      Trinca+23 & \texttt{CAT} & SAM & - & - \\
      \hline
      Xu+16 & \texttt{ENZO} & AMR & 4.3 & 0.21 \\
      Jaacks+19 & \texttt{GIZMO} & MFM & 4 & 0.29 \\
      Liu+20 & \texttt{GIZMO} & MFM & 4 & 0.36 \\
      Sarmento+22 & \texttt{RAMSES-RT} & AMR & 3 & 0.12 \\
      Johnson+13 & \texttt{Gadget-2} & SPH & 2.84 & 0.04 \\
      Visbal+20 & - & SAM & 2.01 & 0.05 \\
      Skinner+20 & \texttt{ENZO} & AMR & 0.67 & 0.01 \\
      Maio+16 & \texttt{Gadget-3} & SPH & 0.5 & 0.04 \\
    \end{tabular}
    \label{tab:PopIII_models}
\end{table}

A comparison of our predictions with small-scale/high-mass-resolution simulations ($L \lesssim 4$~cMpc, Figure~\ref{fig:SFRD_pop_av+modSmallScale}) provides interesting clues on the contribution of well resolved populations of star-forming mini-halos and proper radiative feedback. Pop\ III star formation in small structures follows a quasi-flat evolution in the redshift range $13 \leq z \leq 19$, with values differing by more than one order of magnitude across model predictions. At $z \leq 13$, most of the models predict a flattening or a gentle decline of Pop\ III star formation, and the different behaviour might be ascribed to differences in the implemented hydrodynamical and/or RT scheme, as well as metal mixing/spreading. \citet{Jaacks_2019} and \citet{Sarmento_Scannapieco_2022} are notable exceptions to these trends. Despite the relatively small simulated volume, \citet{Jaacks_2019} find a Pop\ III SFRD  higher than our predictions down to $z \sim 7$, probably because of the relatively inefficient radiative/mechanical feedback model implemented in their simulation\footnote{These aspects were extensively commented in the original paper, and were addressed in later simulations by \citet{Liu_Bromm_2020_1, Liu_Bromm_2020_2}, where they reduced the efficiency of star formation, added SN-driven winds from Pop\ II, and enhanced radiative feedback through a local LW contribution from stellar sources. Indeed, $\Psi_{\rm III}(z)$ predicted by \citet{Liu_Bromm_2020_2} appears substantially reduced, and closer to other small-scale model predictions.}. Conversely, \citet{Sarmento_Scannapieco_2022} find a trend similar to our own down to $z \sim 11 -12$, with a higher level of flattening at lower redshifts and a gentle decline below $z \sim 8$, although they seem to be finding slightly lower values of the Pop\ III SFRD than our model at all redshifts.

A good level of agreement, especially at $z \sim 10 - 14$, is found between our simulations and e.g. the FiBY simulations \citep{Johnson_2013}, notwithstanding the absence of LW feedback in \texttt{dustyGadget}. This can be understood by reminding that LW radiation has an effect on H$_2$ star-forming environments (i.e. mini-halos), unresolved by mass in our simulations. As a result, \texttt{dustyGadget} finds that below $z \simeq 14 - 15$, our predictions of $\Psi_{\rm III}$ are not dramatically different from those of models accounting for LW-suppressed star formation in mini-halos. However, we emphasize again that this also leads to an underestimation of Pop\ III stars at very high redshifts ($z \gtrsim 14 - 15$), where the LW is not yet very efficient, and mini-halos provide the dominant contribution to Pop\ III star formation. 

The study of \citet{Xu_2016_XRB} deserves a particular mention because it provides an interesting comparison between different environments selected from the Reinassance simulation. The authors adopt a zoom-in technique to re-simulate small regions of interest ($\sim 4.3 \; \si{cMpc}h^{-1}$) with increased resolution from a bigger, low-resolution cube of $\sim 28 \; \si{cMpc}h^{-1}$. The chosen sub-regions exhibit different values of the density contrast, and are described as ``Normal'' ($\langle \delta \rangle \simeq 0.09$), ``Void'' ($\langle \delta \rangle \simeq -0.26$) and ``Rarepeak'' ($\langle \delta \rangle \simeq 0.68$). Their $\Psi_{\rm III}(z)$ are shown in the right panel of Figure \ref{fig:SFRD_pop_av+modSmallScale} as gold lines, to visualize the expected spread among different environments. At $11 \lesssim z \lesssim 15$, \texttt{dustyGadget} predictions are consistent with the results of the ``Normal'' region (gold, solid line).

The large differences in both the amplitude and redshift evolution of $\Psi_{\rm III}(z)$ found by small-scale simulations indicate that (i) the estimates of $\Psi_{\rm III}(z)$ are heavily impacted by the bias introduced by the specific scale, (ii) the capability to properly resolve star-forming mini-halos is required to correctly estimate $\Psi_{\rm III}(z)$ at $z \geq 13$, and (iii) the implementation of detailed radiative/chemical feedback is crucial to determine the late-time evolution of Pop\ III star-forming environments and the Pop\ III/II transition. Models differing in both chemical and RT scheme can even predict a sudden and fast suppression of Pop\ III star formation at $z \sim 9$, with extremely low values ($\Psi_{\rm III} \leq 10^{-6.2}$ in $9 < z < 13$), despite their mass resolution \citep{Maio_2016}.

Local radiative feedback, in particular, may have strong implications for gas evolution in star-forming regions. The first three-dimensional radiative simulations studying pristine-gas collapse and the role of UV radiation have been presented in \citet{Machacek_2001}, \citet{Yoshida_2003}, \citet{OShea_2004}, \citet{OShea_Norman_2008}, \citet{Wise_2012_radPressure}, \citet{Wise_2014} and \citet{Regan_2016, Regan_2017}. They show that, depending on the radiative prescriptions adopted, different outcomes for gas collapse are possible and feedback effects are dominant actors during structure formation. Photo-ionization feedback is also discussed in \cite{Kannan_2014}, while LW radiation, which is crucial for star formation quenching and black-hole birth, is instead discussed by e.g. \cite{Habouzit_2016}, \citet{Maio_2019} and \citet{Latif_2021}. In general, their findings suggest that feedback prescriptions are among the main causes of differences in the final results and the formation of Pop\ III stars is tightly linked or even alternative to the birth of massive black holes. In practice, strong radiative fields can locally heat the medium, dissociate $\mathrm{H}_2$ and inhibit cooling and star formation. This process, when happening in unpolluted gas, could lead to the formation of a direct-collapse black hole instead of a Pop\ III star.

We note that the chosen Pop III IMF and critical metallicity $Z_\mathrm{crit}$ can also play a role on the Pop\ III/II transition (see e.g. the discussion in \citealt{Maio_2010}). In the present work, we do not explore the impact of a different Pop III IMFs and/or the critical metallicity on our results, but note that different assumptions have been made by all the models collected here. \citet{Pallottini_2014} and \citet{Maio_2016} provided results using different IMFs: for \citet{Pallottini_2014}, we only display their fiducial model, i.e. a Larson-Salpeter IMF \citep{Larson_1998} in the range [0.1, 100]~\si{M_\odot} (``SALP'' case); for \citet{Maio_2016}, we show the ``TH-1e5K'' case, i.e. a top-heavy IMF over [100, 500]~\si{M_\odot} with slope -2.35, same as ours. Note that while the TH-1e5K case in \citet{Maio_2016} was found to differ by up to 2~dex from the SL-1e4K case (i.e. a Salpeter IMF over the range [0.1, 100]~\si{M_\odot})\footnote{The different behaviour might be ascribed both to different lifetimes and different SN/radiative feedback resulting from Pop\ III stars with different masses. For example, a more powerful spectral energy distribution (i.e. a black body with effective temperature $T_\mathrm{eff} \sim 10^5$~K) is assumed in \citet{Maio_2016} for Pop\ III sources with a top-heavy IMF, while more typical $T_\mathrm{eff} \sim  10^4 - 4 \times 10^4$~K sources were considered for the standard-Salpeter case.}, rather surprisingly, \citet{Pallottini_2014} found essentially no difference between their fiducial model and the model assuming a top-heavy IMF in the range [100, 500]~\si{M_\odot} (``PISN'' case). As for the critical metallicity, most models have assumed $Z_\mathrm{crit} = 10^{-4} ~ \si{Z_\odot}$ (similarly to us), although a lower metallicity threshold is assumed e.g. in \citet{Sarmento_2018} and \citet{Sarmento_Scannapieco_2022} ($Z_\mathrm{crit} = 10^{-5} ~ \si{Z_\odot}$), and in \citet{Skinner_Wise_2020} ($Z_\mathrm{crit} = 5 \times 10^{-6} ~ \si{Z_\odot}$).

\end{appendix}

\label{lastpage}
\end{document}